%% file: main.tex
\pdfoutput=1
\documentclass[12pt,a4paper]{article}

\usepackage{ifthen} 
\newboolean{pdflatex}
\setboolean{pdflatex}{true} 

\newboolean{articletitles}
\setboolean{articletitles}{true} 

\newboolean{uprightparticles}
\setboolean{uprightparticles}{false} 

\def\paperauthors{LHCb collaboration} 
\def\paperasciititle{Model-independent measurement of the transversity amplitudes of the B0->Kst0mumu decay} 
\def\papertitle{Model-independent measurement of the transversity amplitudes of the \BdToKstmm decay} 
\def\paperkeywords{{High Energy Physics}, {LHCb}} 
\def\papercopyright{\the\year\ CERN for the benefit of the LHCb collaboration} 
\def\paperlicence{CC BY 4.0 licence}
\def\paperlicenceurl{https://creativecommons.org/licenses/by/4.0/}

\def\qsqregion{$1.1<$~\qsq~$<8.0$~\gevgevcccc}
\def\Kpmumu{$\Kp\pim\mup\mun$ }
\def\mydecay{\decay{\Bz}{\Kstarz(\to\Kp\pim)\mumu}}

\newif\ifEnableSectionTOCLinks
\EnableSectionTOCLinksfalse 

\input{LHCb/preamble}
\usepackage{longtable} 
\usepackage{multirow}
\usepackage{subcaption}
\usepackage{booktabs}

\begin{document}

\renewcommand{\thefootnote}{\fnsymbol{footnote}}
\setcounter{footnote}{1}

\input{title-LHCb-PAPER}


\renewcommand{\thefootnote}{\arabic{footnote}}
\setcounter{footnote}{0}


\cleardoublepage


\pagestyle{plain} 
\setcounter{page}{1}
\pagenumbering{arabic}


\input{body}

\input{LHCb/acknowledgements}




\addcontentsline{toc}{section}{References}
\bibliographystyle{LHCb/LHCb}
\bibliography{main,LHCb/standard,LHCb/LHCb-PAPER,LHCb/LHCb-CONF,LHCb/LHCb-DP,LHCb/LHCb-TDR,LHCb/LHCb-PUB}

\newpage
\input{Authorship_LHCb-PAPER-2026-023}

\end{document}

%% file: LHCb/preamble.tex
\usepackage[top=1in, bottom=1.25in, left=1in, right=1in]{geometry}

\usepackage{microtype}
\usepackage{lineno}  
\usepackage{xspace} 
\usepackage{caption} 

\usepackage{graphicx}  
\usepackage{color}
\usepackage{colortbl}
\graphicspath{{./figs/}} 

\usepackage{amsmath} 
\usepackage{amssymb}
\usepackage{amsfonts}
\usepackage{upgreek} 

\usepackage[normalem]{ulem} 

\newcommand*\patchAmsMathEnvironmentForLineno[1]{%
\expandafter\let\csname old#1\expandafter\endcsname\csname #1\endcsname
\expandafter\let\csname oldend#1\expandafter\endcsname\csname
end#1\endcsname
 \renewenvironment{#1}%
   {\linenomath\csname old#1\endcsname}%
   {\csname oldend#1\endcsname\endlinenomath}%
}
\newcommand*\patchBothAmsMathEnvironmentsForLineno[1]{%
  \patchAmsMathEnvironmentForLineno{#1}%
  \patchAmsMathEnvironmentForLineno{#1*}%
}
\AtBeginDocument{%
\patchBothAmsMathEnvironmentsForLineno{equation}%
\patchBothAmsMathEnvironmentsForLineno{align}%
\patchBothAmsMathEnvironmentsForLineno{flalign}%
\patchBothAmsMathEnvironmentsForLineno{alignat}%
\patchBothAmsMathEnvironmentsForLineno{gather}%
\patchBothAmsMathEnvironmentsForLineno{multline}%
\patchBothAmsMathEnvironmentsForLineno{eqnarray}%
}

\usepackage[pdftex,
            pdfauthor={\paperauthors},
            pdftitle={\paperasciititle},
            pdfkeywords={\paperkeywords}]{hyperref}
\usepackage{hyperxmp}
\hypersetup{
    pdfcopyright={Copyright (C) \papercopyright},
    pdflicenseurl={\paperlicenceurl}
}

\usepackage[colorinlistoftodos,textsize=scriptsize]{todonotes}

\usepackage[bottom,flushmargin,hang,multiple]{footmisc}

\usepackage[all]{hypcap} 

\input{LHCb/lhcb-symbols-def} 

\hypersetup{
  colorlinks   = true, 
  urlcolor     = blue, 
  linkcolor    = blue, 
  citecolor    = red   
}

\ifEnableSectionTOCLinks
    \usepackage[explicit]{titlesec} 
    
    \let\oldcontentsline\contentsline
    \renewcommand

    \titleformat{\section}{\normalfont\Large\bf}{\hyperlink{tocsection.\thesection}{{\thesection} \parbox[t]{\dimexpr\textwidth-1pc}{#1}}}{1pc}{}

    \titleformat{\subsection}{\normalfont\bf}{\hyperlink{tocsubsection.\thesubsection}{{\thesubsection} \parbox[t]{\dimexpr\textwidth-1pc}{#1}}}{1pc}{}

    \titleformat{name=\section,numberless}[display]{}{}{0pt}{\normalfont\Huge\bfseries #1}
\fi

\usepackage{cite} 
\usepackage{LHCb/mciteplus}

%% file: LHCb/lhcb-symbols-def.tex
\usepackage{xspace} 
\usepackage{upgreek}

\def\lhcb   {\mbox{LHCb}\xspace}

\def\MagUp {\mbox{\em Mag\kern -0.05em Up}\xspace}

\ifthenelse{\boolean{uprightparticles}}%
{

 \def\Pmu         {\ensuremath{\upmu}\xspace}                 
 \def\Pnu         {\ensuremath{\upnu}\xspace}                 
                  
 \def\Ppi         {\ensuremath{\uppi}\xspace}

 \def\Ppsi        {\ensuremath{\uppsi}\xspace}

 \def\PDelta      {\ensuremath{\Delta}\xspace}                 
 \def\PXi         {\ensuremath{\Xi}\xspace}                 
 \def\PLambda     {\ensuremath{\Lambda}\xspace}                 
 \def\PSigma      {\ensuremath{\Sigma}\xspace}                 
 \def\POmega      {\ensuremath{\Omega}\xspace}                 
 \def\PUpsilon    {\ensuremath{\Upsilon}\xspace}
 \let\oldPi\Pi
 \def\PPi         {\ensuremath{\oldPi}\xspace}

 \def\PB      {\ensuremath{\mathrm{B}}\xspace}                 
 \def\PD      {\ensuremath{\mathrm{D}}\xspace}                 
 \def\PJ      {\ensuremath{\mathrm{J}}\xspace}                 
 \def\PK      {\ensuremath{\mathrm{K}}\xspace}                 
 \def\Pb      {\ensuremath{\mathrm{b}}\xspace}                 
 \def\Pc      {\ensuremath{\mathrm{c}}\xspace}

 \def\Pp      {\ensuremath{\mathrm{p}}\xspace}                 

 \def\Ps      {\ensuremath{\mathrm{s}}\xspace}

 \def\thebaroffset{0.0em}
}
{

 \def\Pmu         {\ensuremath{\mu}\xspace}                 
 \def\Pnu         {\ensuremath{\nu}\xspace}                 
                  
 \def\Ppi         {\ensuremath{\pi}\xspace}

 \def\Ppsi        {\ensuremath{\psi}\xspace}                 
                  
 \mathchardef\PDelta="7101
 \mathchardef\PXi="7104
 \mathchardef\PLambda="7103
 \mathchardef\PSigma="7106
 \mathchardef\POmega="710A
 \mathchardef\PUpsilon="7107
 \mathchardef\PPi="7105
 \def\PB      {\ensuremath{B}\xspace}                 
 \def\PD      {\ensuremath{D}\xspace}                 
 \def\PJ      {\ensuremath{J}\xspace}                 
 \def\PK      {\ensuremath{K}\xspace}                 
 \def\Pb      {\ensuremath{b}\xspace}                 
 \def\Pc      {\ensuremath{c}\xspace}

 \def\Pp      {\ensuremath{p}\xspace}                 

 \def\Ps      {\ensuremath{s}\xspace}

 \def\thebaroffset{0.18em}
}
\newcommand{\offsetoverline}[2][\thebaroffset]{\kern #1\overline{\kern -#1 #2}}%

\makeatletter
\ifcase \@ptsize \relax
  \newcommand{\miniscule}{\@setfontsize\miniscule{4}{5}}
\or
  \newcommand{\miniscule}{\@setfontsize\miniscule{5}{6}}
\or
  \newcommand{\miniscule}{\@setfontsize\miniscule{5}{6}}
\fi
\makeatother

\DeclareRobustCommand{\optbar}[1]{\shortstack{{\miniscule (\rule[.5ex]{1.25em}{.18mm})}
  \\ [-.7ex] $#1$}}

\def\mup        {{\ensuremath{\Pmu^+}}\xspace}
\def\mun        {{\ensuremath{\Pmu^-}}\xspace} 

\def\mumu       {{\ensuremath{\Pmu^+\Pmu^-}}\xspace}

\def\neu        {{\ensuremath{\Pnu}}\xspace}
\def\neub       {{\ensuremath{\overline{\Pnu}}}\xspace}
\def\neum       {{\ensuremath{\neu_\mu}}\xspace}
\def\neumb      {{\ensuremath{\neub_\mu}}\xspace}

\def\squark    {{\ensuremath{\Ps}}\xspace}

\def\cquark    {{\ensuremath{\Pc}}\xspace}

\def\bquark    {{\ensuremath{\Pb}}\xspace}

\def\pion   {{\ensuremath{\Ppi}}\xspace}
\def\piz    {{\ensuremath{\pion^0}}\xspace}
\def\pip    {{\ensuremath{\pion^+}}\xspace}
\def\pim    {{\ensuremath{\pion^-}}\xspace}

\def\kaon    {{\ensuremath{\PK}}\xspace}
\def\Kbar    {{\ensuremath{\offsetoverline{\PK}}}\xspace}

\def\KorKbar {\kern \thebaroffset\optbar{\kern -\thebaroffset \PK}{}\xspace}

\def\Kp      {{\ensuremath{\kaon^+}}\xspace}
\def\Km      {{\ensuremath{\kaon^-}}\xspace}

\def\KS      {{\ensuremath{\kaon^0_{\mathrm{S}}}}\xspace}

\def\Kstarz  {{\ensuremath{\kaon^{*0}}}\xspace}
\def\Kstarzb {{\ensuremath{\Kbar{}^{*0}}}\xspace}

\def\Kstarp  {{\ensuremath{\kaon^{*+}}}\xspace}

\def\Dbar    {{\ensuremath{\offsetoverline{\PD}}}\xspace}
\def\D       {{\ensuremath{\PD}}\xspace}

\def\DorDbar {\kern \thebaroffset\optbar{\kern -\thebaroffset \PD}\xspace}

\def\Dzb     {{\ensuremath{\Dbar{}^0}}\xspace}
\def\Dp      {{\ensuremath{\D^+}}\xspace}
\def\Dm      {{\ensuremath{\D^-}}\xspace}

\def\DpDm    {\ensuremath{\Dp {\kern -0.16em \Dm}}\xspace}

\def\B       {{\ensuremath{\PB}}\xspace}
\def\Bbar    {{\ensuremath{\offsetoverline{\PB}}}\xspace}

\def\BorBbar {\kern \thebaroffset\optbar{\kern -\thebaroffset \PB}\xspace}
\def\Bz      {{\ensuremath{\B^0}}\xspace}
\def\Bzb     {{\ensuremath{\Bbar{}^0}}\xspace}
\def\Bd      {{\ensuremath{\B^0}}\xspace}
\def\Bdb     {{\ensuremath{\Bbar{}^0}}\xspace}
\def\BdorBdbar {\kern \thebaroffset\optbar{\kern -\thebaroffset \Bd}\xspace}
\def\Bu      {{\ensuremath{\B^+}}\xspace}

\def\Bp      {{\ensuremath{\Bu}}\xspace}

\def\Bs      {{\ensuremath{\B^0_\squark}}\xspace}

\def\BsorBsbar {\kern \thebaroffset\optbar{\kern -\thebaroffset \Bs}\xspace}

\def\jpsi     {{\ensuremath{{\PJ\mskip -3mu/\mskip -2mu\Ppsi}}}\xspace}
\def\psitwos  {{\ensuremath{\Ppsi{(2S)}}}\xspace}

\def\Y#1S{\ensuremath{\PUpsilon{(#1S)}}\xspace}

\def\proton      {{\ensuremath{\Pp}}\xspace}

\def\Lz          {{\ensuremath{\PLambda}}\xspace}

\def\LorLbar     {\kern \thebaroffset\optbar{\kern -\thebaroffset \PLambda}\xspace}

\def\Lb           {{\ensuremath{\Lz^0_\bquark}}\xspace}

\newcommand{\decay}[2]{\ensuremath{\mathinner{#1\!\to #2}}\xspace}

\def\to                 {\ensuremath{\rightarrow}\xspace}

\def\qsq       {{\ensuremath{q^2}}\xspace}

\def\CP                {{\ensuremath{C\!P}}\xspace}

\def\BdToKstmm    {\decay{\Bd}{\Kstarz\mup\mun}}

\def\BdToJPsiKst  {\decay{\Bd}{\jpsi\Kstarz}}

\def\bsll     {\decay{\bquark}{\squark \ell^+ \ell^-}}

\def\AT#1     {\ensuremath{A_{\mathrm{T}}^{#1}}\xspace}           

\def\ctl       {\ensuremath{\cos{\theta_\ell}}\xspace}
\def\ctk       {\ensuremath{\cos{\theta_K}}\xspace}

\def\C#1      {\ensuremath{\mathcal{C}_{#1}}\xspace}                       
\def\Cp#1     {\ensuremath{\mathcal{C}_{#1}^{'}}\xspace}                    
\def\Ceff#1   {\ensuremath{\mathcal{C}_{#1}^{\mathrm{(eff)}}}\xspace}        
\def\Cpeff#1  {\ensuremath{\mathcal{C}_{#1}^{'\mathrm{(eff)}}}\xspace}       
\def\Ope#1    {\ensuremath{\mathcal{O}_{#1}}\xspace}                       
\def\Opep#1   {\ensuremath{\mathcal{O}_{#1}^{'}}\xspace}                    

\newcommand{\nospaceunit}[1]{\ensuremath{\text{#1}}}       
\newcommand{\aunit}[1]{\ensuremath{\text{\,#1}}}       

\newcommand{\tev}{\aunit{Te\kern -0.1em V}\xspace}
\newcommand{\gev}{\aunit{Ge\kern -0.1em V}\xspace}
\newcommand{\mev}{\aunit{Me\kern -0.1em V}\xspace}
\newcommand{\kev}{\aunit{ke\kern -0.1em V}\xspace}
\newcommand{\ev}{\aunit{e\kern -0.1em V}\xspace}
 
\newcommand{\mevc}{\ensuremath{\aunit{Me\kern -0.1em V\!/}c}\xspace}
\newcommand{\gevc}{\ensuremath{\aunit{Ge\kern -0.1em V\!/}c}\xspace}
\newcommand{\mevcc}{\ensuremath{\aunit{Me\kern -0.1em V\!/}c^2}\xspace}
\newcommand{\gevcc}{\ensuremath{\aunit{Ge\kern -0.1em V\!/}c^2}\xspace}
\newcommand{\gevgevcccc}{\ensuremath{\gev^2\!/c^4}\xspace} 

\def\mum  {\ensuremath{\,\upmu\nospaceunit{m}}\xspace}

\def\fb   {\ensuremath{\aunit{fb}}\xspace}
\def\invfb   {\ensuremath{\fb^{-1}}\xspace}

\def\deriv {\ensuremath{\mathrm{d}}}

\def\gsim{{~\raise.15em\hbox{$>$}\kern-.85em
          \lower.35em\hbox{$\sim$}~}\xspace}
\def\lsim{{~\raise.15em\hbox{$<$}\kern-.85em
          \lower.35em\hbox{$\sim$}~}\xspace}

\def\sPlot{\mbox{\em sPlot}\xspace}

\def\pt         {\ensuremath{p_{\mathrm{T}}}\xspace}

\def\ptot       {\ensuremath{p}\xspace}

\def\evtgen     {\mbox{\textsc{EvtGen}}\xspace}

\def\geant      {\mbox{\textsc{Geant4}}\xspace}

\def\photos     {\mbox{\textsc{Photos}}\xspace}

\def\pythia     {\mbox{\textsc{Pythia}}\xspace}

\def\tell1  {TELL1\xspace}
\def\ukl1   {UKL1\xspace}

\newcommand{\lhcborcid}[1]{\href{https://orcid.org/#1}{\hspace*{0.1em}\raisebox{-0.45ex}{\includegraphics[width=1em]{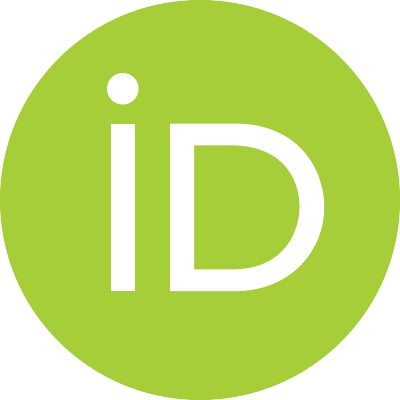}}}}


%% file: title-LHCb-PAPER.tex

\begin{titlepage}
\pagenumbering{roman}

\vspace*{-1.5cm}
\centerline{\large EUROPEAN ORGANIZATION FOR NUCLEAR RESEARCH (CERN)}
\vspace*{1.5cm}
\noindent
\begin{tabular*}{\linewidth}{lc@{\extracolsep{\fill}}r@{\extracolsep{0pt}}}
\ifthenelse{\boolean{pdflatex}}
{\vspace*{-1.5cm}\mbox{\!\!\!\includegraphics[width=.14\textwidth]{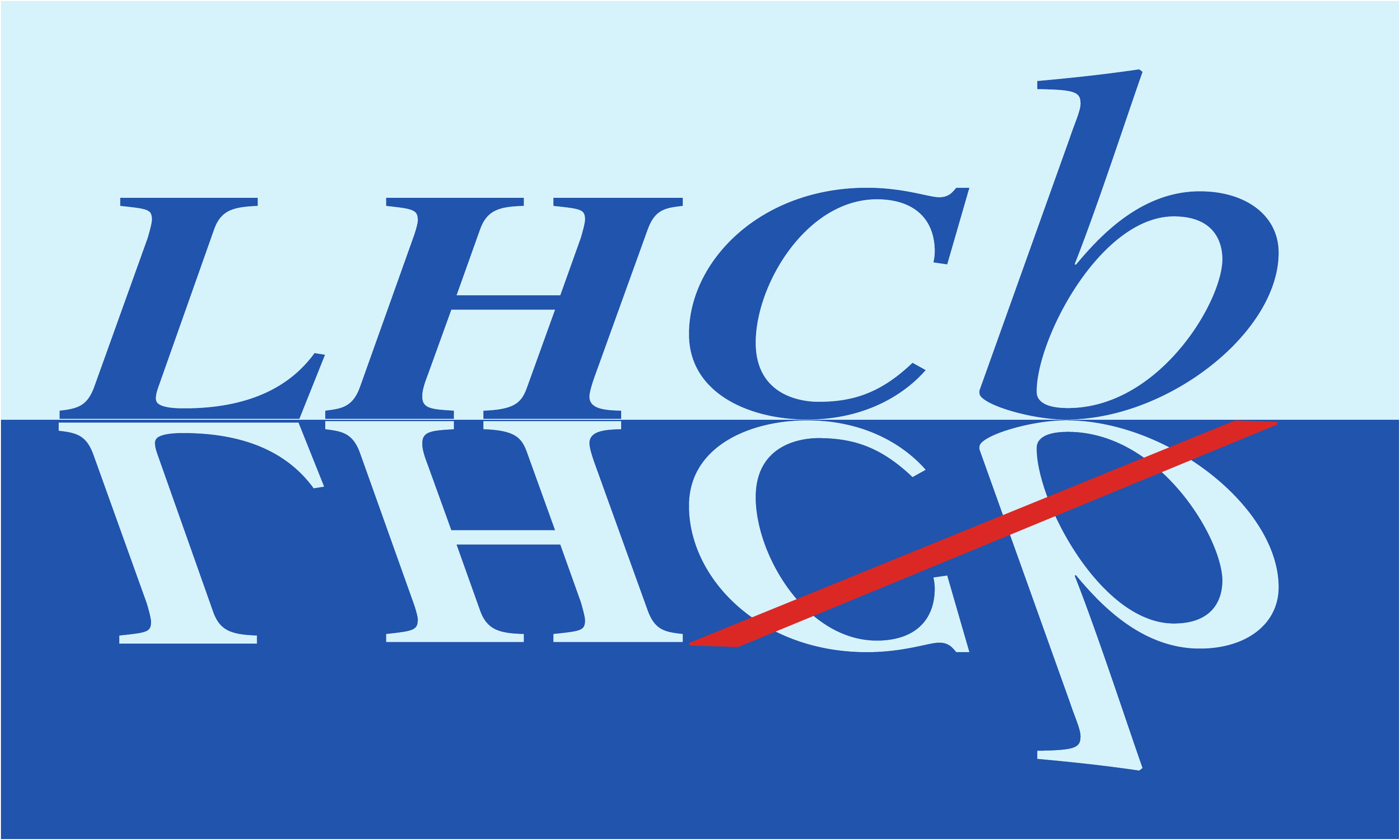}} & &}%
{\vspace*{-1.2cm}\mbox{\!\!\!\includegraphics[width=.12\textwidth]{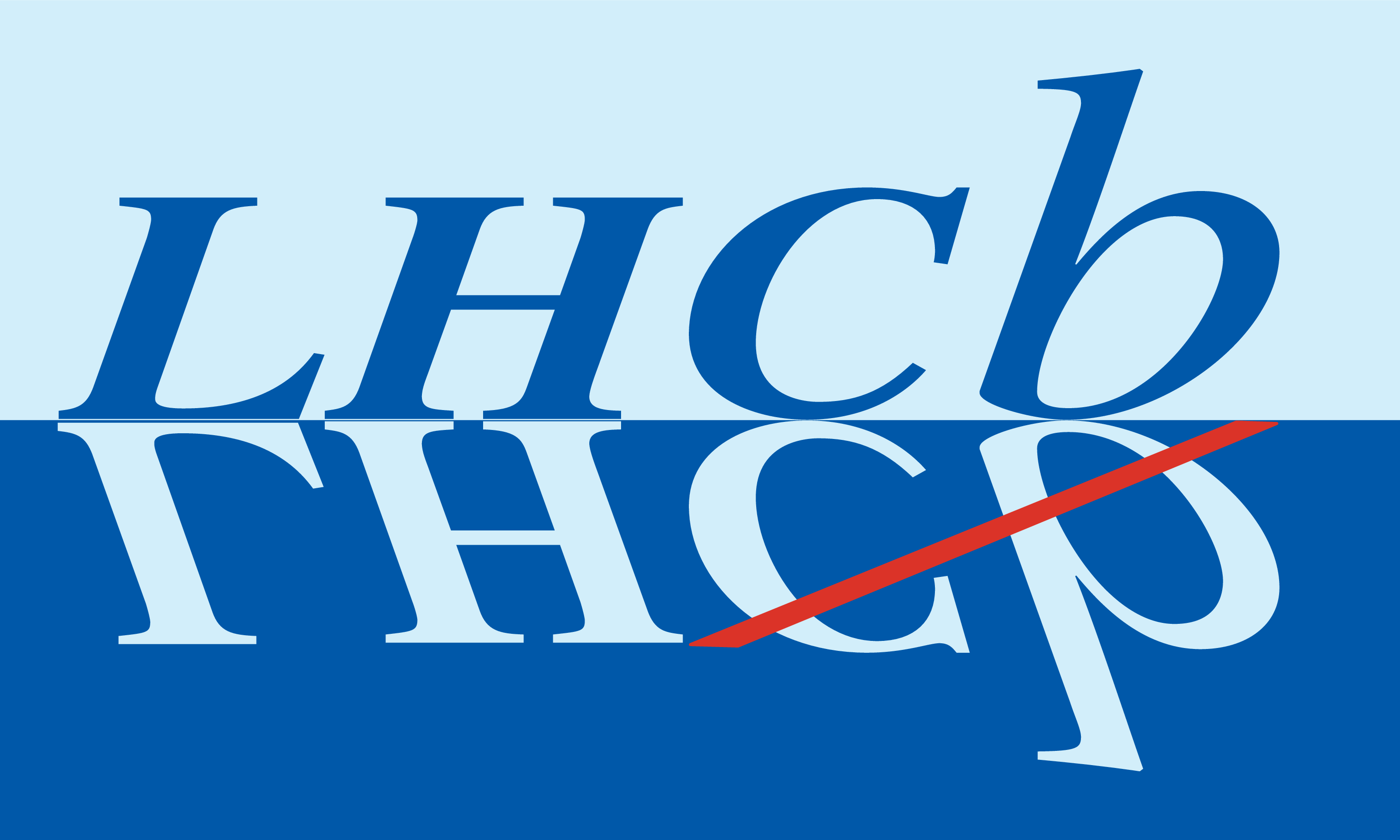}} & &}%
\\
 & & CERN-EP-2026-226 \\  
 & & LHCb-PAPER-2026-023 \\  
 & & 12 August 2026 \\ 
 & & \\
\end{tabular*}

\vspace*{4.0cm}

{\normalfont\bfseries\boldmath\huge
\begin{center}
  \papertitle 
\end{center}
}

\vspace*{2.0cm}

\begin{center}
\paperauthors\footnote{Authors are listed at the end of this paper.}
\end{center}

\vspace{\fill}

\begin{abstract}
  \noindent
  An analysis of the decay amplitudes of \mydecay is presented, using proton-proton collision data recorded by the \lhcb experiment at centre-of-mass energies of 7, 8, and 13\tev, corresponding to an integrated luminosity of 8.4\invfb. The amplitudes are constructed from Legendre polynomials in the $\mumu$ invariant mass squared region \qsqregion. \CP-averaged observables are obtained from the amplitudes. Some of these observables present deviations with respect to the Standard Model, which can be interpreted as shifts in the effective Wilson coefficients. This model-independent approach enables tests of theoretical predictions that can help disentangle hadronic effects from potential contributions from physics beyond the Standard Model. This allows flexibility in the choice of \qsq binning for global analyses. Depending on the binning scheme, the deviation of the Wilson coefficient $C_9$ from its Standard Model expectation varies from $4.3\sigma$ to $4.8\sigma$.
\end{abstract}

\vspace*{2.0cm}

\begin{center}
  Submitted to
  JHEP 
\end{center}

\vspace{\fill}

{\footnotesize 
\centerline{\copyright~\papercopyright. \href{\paperlicenceurl}{\paperlicence}.}}
\vspace*{2mm}

\end{titlepage}


\newpage
\setcounter{page}{2}
\mbox{~}
%
%
%
%

%% file: body.tex
\section{Introduction}
\label{sec:Introduction}

Flavour-changing neutral currents are suppressed in the Standard Model (SM) and are thus sensitive to possible additional contributions stemming from New Physics (NP) processes.
Recent measurements of \bsll transitions have revealed tensions with SM predictions. These include measurements of branching fractions~\cite{LHCb-PAPER-2014-006,LHCb-PAPER-2015-023,LHCb-PAPER-2016-012,LHCb-PAPER-2021-014,BELLE:2019xld,Belle:2019oag} and angular observables~\cite{LHCb-PAPER-2020-002,LHCb-PAPER-2021-022,LHCb-PAPER-2020-041,ATLAS:2018gqc,CMS:2015bcy,CMS:2020oqb}, which exhibit coherent inconsistencies with respect to the SM in several channels. These tensions are collectively known as flavour anomalies and have been of significant interest in recent years \cite{Capdevila:2023yhq}.

One such example is the process \decay{\Bz}{\Kstarz\mu^+\mu^-}. A measurement of branching fraction and angular observables, in bins of $\mumu$ invariant mass squared, \qsq, has shown discrepancies at the level of $3.6$ to $4.1$ standard deviations \cite{LHCb-PAPER-2025-041}. Global fits also show that tensions in \bsll measurements with respect to SM predictions~\cite{Gubernari:2022hxn, hurth2023banomaliespostrk, Capdevila:2023yhq} are observed up to the level of six standard deviations.
Recent theoretical studies~\cite{Isidori_2025, bordone2024shortvslongdistancephysics, bordone2026disentanglingshortvslongdistance} have shown that SM hadronic effects and NP contributions have different \qsq behaviours, motivating a better understanding of the \qsq-dependence of the angular observables. Thus, in order to attain increased sensitivity to these effects, an analysis where the observables are measured unbinned in \qsq is of great interest.
In addition, measuring the decay amplitudes provides a direct parameterisation of the angular distribution from which the full set of angular observables can be constructed \cite{Matias_2012}.

The LHCb collaboration has performed analyses of \BdToKstmm transitions using model-dependent descriptions of the \qsq-dependence of this decay with the aim of separating out long- and short-distance contributions directly from the data~\cite{LHCb-PAPER-2024-011,LHCb-PAPER-2023-033}. As the accuracy of long-distance contributions in such models remains a subject of theoretical debate, a model-independent description of the decay amplitudes that exploits the added precision achievable with a \qsq-unbinned approach, is essential.

As outlined in Ref.~\cite{Egede_2015}, the \BdToKstmm amplitudes can be described through a \qsq-dependent empirical ansatz, without explicitly separating out the short- and long-distance contributions or relying on external theory predictions. 
Using such an amplitude parametrisation allows for the measurement of angular observables as continuous functions of \qsq, providing a way to test whether or not the \bsll flavour anomalies can be ascribed to NP or SM hadronic effects~\cite{bordone2024shortvslongdistancephysics}.

This paper describes an unbinned amplitude analysis of the decay \BdToKstmm in the \qsq range \qsqregion, where the \Kstarz is reconstructed through the \Kp \pim final state, and $\Kstarz$ denotes the $K^*(892)^0$ meson.\footnote{Charge-conjugate processes are implied unless otherwise stated.} The \qsq region is chosen to be between the $\phi$ and \jpsi resonances since in these regions the amplitudes in \qsq vary rapidly and may not be correctly modelled.
This analysis is performed on a combined data sample corresponding to an integrated luminosity of $8.4$~fb$^{-1}$ of proton-proton (\proton\proton) collisions collected by the LHCb experiment at centre-of-mass energies of 7, 8, and 13 \tev during the 2011--2012 and 2016--2018 data taking periods.
In this study, Legendre polynomials are used to describe the \qsq-dependence of the amplitudes. This choice is motivated by the orthogonality of Legendre polynomials and the reduced correlations between amplitude components, which provide improved fit stability as compared to previous empirical parametrisations~\cite{Egede_2015,LHCb-PAPER-2015-051}. The measurement is performed on the combined \Bz and \Bzb datasets, thus the amplitudes are used to construct the \CP-averaged observables. This analysis uses the same triggers, first stage selection, and initial offline processing as Ref.~\cite{LHCb-PAPER-2025-041}. Furthermore, the data-simulation corrections used for this analysis and the corrections used to compute systematic uncertainties correspond to the nominal sets of corrections used in Ref.~\cite{LHCb-PAPER-2025-041}.
This paper is organised as follows. The differential decay rate of \BdToKstmm is described in Sec.~\ref{sec:angular}. The LHCb detector is described in Sec.~\ref{sec:experiment}. The signal selection is outlined in Sec.~\ref{sec:Selection} and its acceptance is described in Sec.~\ref{sec:acceptance}. Details of the fitting procedure are given in Sec.~\ref{sec:fit}, and systematic uncertainties are described in Sec.~\ref{sec:systematics}. The results of the analysis are presented in Sec.~\ref{sec:results} with concluding remarks in Sec.~\ref{sec:conclusions}.

\section{Differential decay rate}
\label{sec:angular}

The decay \BdToKstmm, followed by \decay{\Kstarz}{\Kp\pim}, is fully described by five parameters: the squared invariant mass of the $\Kp\pim$ system, $m_{\kaon\pion}^{2}$; $\qsq$; and three angles $\theta_\ell$, $\theta_K$, and $\phi$. As described in Refs.~\cite{Altmannshofer_2009, PhysRevD.93.054008}, $\theta_\ell$ is defined as the angle between the \mup (\mun) direction in the $\mumu$ rest frame and the direction of the $\mumu$ system in the \Bd (\Bdb) rest frame. The angle $\theta_K$ is the angle between the direction of the $K^+$ ($K^-$) meson in the \Kstarz (\Kstarzb) rest frame and the direction of the \Kstarz (\Kstarzb) in the \Bd (\Bdb) rest frame. The angle $\phi$ is the angle between the plane containing the \mup and \mun leptons and the plane containing the $K^+$ ($K^-$) and $\pi^-$ ($\pi^+$) mesons, in the \Bd (\Bdb) rest frame.

The differential decay rate, integrated over $m_{\kaon\pion}^{2}$, is written as a function of the quantities \ctl, \ctk, and $\phi$, with the angular distribution for this decay in the case of a P-wave $K\pi$ system given by~\cite{Altmannshofer_2009}

\begin{equation}
\begin{aligned}
\frac{\deriv\Gamma_{\textrm{P}}(\BdToKstmm)}{\deriv\ctl{}~\deriv\ctk{}~\deriv\phi~\deriv\qsq}
=& \frac{9}{32\pi}\big[
J_{1s}\sin^{2}\theta_{K} + J_{1c}\cos^{2}\theta_{K} \\
&\quad + J_{2s}\sin^{2}\theta_{K}\cos2\theta_{\ell}
+ J_{2c}\cos^{2}\theta_{K}\cos2\theta_{\ell} \\
&\quad + J_{3}\sin^{2}\theta_{K}\sin^{2}\theta_{\ell}\cos2\phi
+ J_{4}\sin2\theta_{K}\sin2\theta_{\ell}\cos\phi \\
&\quad + J_{5}\sin2\theta_{K}\sin\theta_{\ell}\cos\phi
+ J_{6s}\sin^{2}\theta_{K}\ctl\\
&\quad + J_{6c}\cos^{2}\theta_{K}\ctl 
+ J_{7}\sin2\theta_{K}\sin\theta_{\ell}\sin\phi \\
&\quad+ J_{8}\sin2\theta_{K}\sin2\theta_{\ell}\sin\phi 
 + J_{9}\sin^{2}\theta_{K}\sin^{2}\theta_{\ell}\sin2\phi
\big],
\end{aligned}
\end{equation}
and for the S-wave (including P-S interference terms)
\begin{equation}
\begin{aligned}
    \frac{\deriv\Gamma_{\textrm{S}}(\BdToKstmm)}{\deriv\ctl{}~\deriv\ctk{}~\deriv\phi~\deriv \qsq} =& \frac{1}{4\pi}\big[  ( \tilde{J}^{c}_{1a} + \tilde{J}^{c}_{2a} \cos 2 \theta_{\ell})  \\
    &\quad + \tilde{J}^{c}_{1b} \ctk + \tilde{J}^{c}_{2b} \cos2\theta_{\ell} \ctk  \\
    &\quad + \tilde{J}_{4} \sin2\theta_{\ell} \sin \theta_{K} \cos \phi + \tilde{J}_{5}  \sin \theta_{\ell} \sin \theta_{K} \cos \phi  \\
    &\quad + \tilde{J}_{7} \sin \theta_{\ell} \sin \theta_{K} \sin \phi + \tilde{J}_{8} \sin 2 \theta_{\ell} \sin \theta_{K} \sin \phi \big].
\end{aligned}
\end{equation}

Assuming no scalar and timelike amplitudes, the coefficients $J_i \equiv J_i (\qsq)$ and $\tilde{J}_i\equiv\tilde{J}_i(\qsq)$ are written in terms of bilinear combinations of eight complex decay amplitudes. Six of these correspond to the different transversity states of the $\Kp \pim$ system in a P-wave configuration denoted as $\mathcal{A}^{\textrm{L},\textrm{R}}_\parallel$, $\mathcal{A}^{\textrm{L},\textrm{R}}_\perp$, and $\mathcal{A}^{\textrm{L},\textrm{R}}_0$. The $\Kp \pim$ system in an S-wave configuration is described by two amplitudes, $\mathcal{A}^{\textrm{L},\textrm{R}}_{00}$. In these expressions, L and R denote left- and right-chiralities of the $\mumu$ system. The kinematic prefactors due to the muon mass are not negligible in this analysis. 

Thus the expressions for $J$ and $\tilde{J}$ are formulated using the amplitudes detailed in Ref.~\cite{Matias_2012} and preserve terms with prefactors of $\beta_\mu$, where $\beta_\mu \equiv \sqrt{1-\frac{4 m_\mu^2}{\qsq}}$.

Following the approach of~Ref.~\cite{Matias_2012}, and excluding coefficients with explicit prefactors of $\frac{m_\mu^2}{\qsq}$, the P-wave $J$ observables are written as 

\begin{equation}
\begin{aligned}
J_{1s} &= \frac{2+\beta_\mu^2}{4}  \left[ |{\cal A}_{\parallel}^{\rm L}|^{2} + |{\cal A}_{\perp}^{\rm L}|^{2}  + |{\cal A}_{\parallel}^{\rm R}|^{2} + |{\cal A}_{\perp}^{\rm R}|^{2}\right],\\
J_{1c} &= |{\cal A}_{0}^{\rm L}|^{2} +  |{\cal A}_{0}^{\rm R}|^{2},\\
J_{2s} &= \frac{1}{4} \beta_\mu^2 \left[ |{\cal A}_{\parallel}^{\rm L}|^{2} + |{\cal A}_{\perp}^{\rm L}|^{2}  + |{\cal A}_{\parallel}^{\rm R}|^{2} + |{\cal A}_{\perp}^{\rm R}|^{2}\right]\\
J_{2c} &= - \beta_\mu^2 \left[  |{\cal A}_{0}^{\rm L}|^{2} +  |{\cal A}_{0}^{\rm R}|^{2} \right],\\
J_{3} &= \frac{1}{2}  \beta_\mu^2   \left[ |{\cal A}_{\perp}^{\rm L}|^{2} - |{\cal A}_{\parallel}^{\rm L}|^{2}  + |{\cal A}_{\perp}^{\rm R}|^{2} - |{\cal A}_{\parallel}^{\rm R}|^{2} \right],\\
J_{4} &= \sqrt{\frac{1}{2}}   \beta_\mu^2  {\rm Re}({\cal A}_{0}^{\rm L} {\cal A}_{\parallel}^{{\rm L}\ast} + {\cal A}_{0}^{\rm R} {\cal A}_{\parallel}^{{\rm R}\ast}),\\
J_{5} &= \sqrt{2} \beta_\mu {\rm Re}({\cal A}_{0}^{\rm L} {\cal A}_{\perp}^{{\rm L}\ast} - {\cal A}_{0}^{\rm R} {\cal A}_{\perp}^{{\rm R}\ast}),\\
J_{6s} &= 2 \beta_\mu  {\rm Re}({\cal A}_{\parallel}^{\rm L} {\cal A}_{\perp}^{{\rm L}\ast} - {\cal A}_{\parallel}^{\rm R} {\cal A}_{\perp}^{{\rm R}\ast}),\\
J_{6c} &= 0,\\
J_{7} &= \sqrt{2}  \beta_\mu {\rm Im}({\cal A}_{0}^{\rm L} {\cal A}_{\parallel}^{{\rm L}\ast} - {\cal A}_{0}^{\rm R} {\cal A}_{\parallel}^{{\rm R}\ast}),\\
J_{8} &= \sqrt{\frac{1}{2}}  \beta_\mu^2  {\rm Im}({\cal A}_{0}^{\rm L} {\cal A}_{\perp}^{{\rm L}\ast} + {\cal A}_{0}^{\rm R} {\cal A}_{\perp}^{{\rm R}\ast}),\\
J_{9} &= \beta_\mu^2  {\rm Im}({\cal A}_{\parallel}^{{\rm L}\ast} {\cal A}_{\perp}^{\rm L} + {\cal A}_{\parallel}^{{\rm R}\ast} {\cal A}_{\perp}^{\rm R}).\\
\end{aligned}
\end{equation}
One can also write the S-wave amplitudes in terms of their real and imaginary components and the chiralities of the $\mumu$ system. In terms of the amplitudes, the S-wave and interference $\tilde{J}$-terms are written as \cite{Alguer_2021}
\begin{equation}
\begin{aligned}
\tilde{J}^{c}_{1a} &= \frac{3}{8}  \left[ |{\cal A}_{\rm 00}^{\rm L}|^{2} + |{\cal A}_{\rm 00}^{\rm R}|^{2}\right],\\
\tilde{J}^{c}_{2a} &= -\frac{3}{8} \beta_\mu^2 \left[ |{\cal A}_{\rm 00}^{\rm L}|^{2} + |{\cal A}_{\rm 00}^{\rm R}|^{2}\right],\\
\tilde{J}^{c}_{1b} &= \sqrt{\frac{27}{16}}  {\rm Re}({\cal A}_{\rm 00}^{\rm L} {\cal A}_{0}^{{\rm L}\ast} + {\cal A}_{\rm 00}^{\rm R} {\cal A}_{0}^{{\rm R}\ast}),\\
\tilde{J}^{c}_{2b} &= -\sqrt{\frac{27}{16}} \beta_\mu^2 {\rm Re}({\cal A}_{\rm 00}^{\rm L} {\cal A}_{0}^{{\rm L}\ast} + {\cal A}_{\rm 00}^{\rm R} {\cal A}_{0}^{{\rm R}\ast}),\\
\tilde{J}_{4} &= \sqrt{\frac{27}{32}}  \beta_\mu^2 {\rm Re}({\cal A}_{\rm 00}^{\rm L} {\cal A}_{\parallel}^{{\rm L}\ast} + {\cal A}_{\rm 00}^{\rm R} {\cal A}_{\parallel}^{{\rm R}\ast}),\\
\tilde{J}_{5} &= \sqrt{\frac{27}{8}} \beta_\mu  {\rm Re}({\cal A}_{\rm 00}^{\rm L} {\cal A}_{\perp}^{{\rm L}\ast} - {\cal A}_{\rm 00}^{\rm R} {\cal A}_{\perp}^{{\rm R}\ast}),\\
\tilde{J}_{7} &= \sqrt{\frac{27}{8}} \beta_\mu {\rm Im}({\cal A}_{\rm 00}^{\rm L} {\cal A}_{\parallel}^{{\rm L}\ast} - {\cal A}_{\rm 00}^{\rm R} {\cal A}_{\parallel}^{{\rm R}\ast}),\\
\tilde{J}_{8} &= \sqrt{\frac{27}{32}} \beta_\mu^2  {\rm Im}({\cal A}_{\rm 00}^{\rm L} {\cal A}_{\perp}^{{\rm L}\ast} + {\cal A}_{\rm 00}^{\rm R} {\cal A}_{\perp}^{{\rm R}\ast}).
\end{aligned}
\end{equation}

Thus, combining the P-wave, S-wave, and interference, the differential decay rate of \BdToKstmm is given by
\begin{align}
\label{eqn:total}
    \frac{\deriv^4 \Gamma (\BdToKstmm)}{\deriv\ctl{}~\deriv\ctk{}~\deriv\phi~\deriv \qsq} =  \frac{\deriv\Gamma_{\textrm{P}}(\BdToKstmm)}{\deriv\ctl{}~\deriv\ctk{}~\deriv\phi~\deriv \qsq} + \frac{\deriv\Gamma_{\textrm{S}}(\BdToKstmm)}{\deriv\ctl{}~\deriv\ctk{}~\deriv\phi~\deriv \qsq} .
\end{align}
The decay amplitudes are functions of $\mumu$ invariant mass squared, \qsq. The model used for each of the P-wave amplitudes employs a linear combination of four Legendre polynomials
\begin{align}
    \mathcal{A}(\qsq) &= \alpha_0 L_0(\qsq) + \alpha_1 L_1(\qsq) + \alpha_2 L_2(\qsq) + \alpha_3 L_3(\qsq)
\end{align}
where $\alpha_i$ are the amplitude components, $L_i (\qsq)$ are Legendre polynomials of order $i$, and \qsq is mapped to the interval $[-1,1]$, on which the Legendre polynomials are defined. The real and imaginary parts of the amplitude components are measured rather than the magnitude and phase in order to avoid discontinuities in the \qsq-dependence.

The amplitude basis can undergo a rotation by two global phases (one which rotates the left-handed amplitudes and another which rotates the right-handed amplitudes), a rotation about the real and imaginary components, and a rotation that mixes the real and imaginary components, all leaving the decay rate invariant. Due to these continuous symmetries of the P-wave differential decay rate \cite{Matias_2012}, the number of P-wave amplitudes is not equal to the physically observable degrees of freedom. Thus, a basis-fixing condition is applied in order to obtain a unique amplitude basis:
\begin{align}
\label{eqn:ampbasis}
    \textrm{Im}(\mathcal{A}_\perp^R) &= \textrm{Im}(\mathcal{A}_0^L) = \textrm{Re}(\mathcal{A}_0^R) = \textrm{Im}(\mathcal{A}_0^R) = 0.
\end{align}

This amplitude basis is chosen to keep the remaining amplitudes smooth and free of discontinuities, allowing them to be well described by the parameterisation across a wide range of physics models. After fixing the amplitude basis there remain two discrete symmetries $\mathcal{A}_L \leftrightarrow -\mathcal{A}_L$ and $\mathcal{A}_R \leftrightarrow -\mathcal{A}_R$, which also leave the differential decay rate invariant. There are therefore four sets of amplitudes which result in identical differential decay rates.
In this analysis, the convention where  $\textrm{Re}(\mathcal{A}_0^L)$ and $\textrm{Re}(\mathcal{A}_\parallel^R)$ have positive values for $\alpha_0$ is chosen in order to break this degeneracy.

In summary, since there are four parameters for each P-wave amplitude, and there are eight independent real and imaginary parts of the amplitudes, there are 32 parameters used to describe the differential decay rate. These are extracted via an unbinned extended maximum-likelihood fit of the angular distribution in the region \qsqregion. In this fit, the \Kpmumu mass, \qsq, \ctl, \ctk, and $\phi$ are fitted simultaneously. The S-wave amplitude is treated as a nuisance parameter with its left and right, real and imaginary components modelled as a uniform function in \qsq. The impact of this assumption was investigated and found to be negligible compared to the statistical uncertainty since the overall precision on the S-wave is expected to be low because the differential decay rate is integrated over $m_{\kaon\pion}$. From the amplitude components, the amplitudes and angular observables can be calculated. In particular, the $J$ observables normalised to the total decay rate, defined as
\begin{align}
\label{eqn:Ss}
{S_i \equiv J_i \times \left(\frac{\deriv \Gamma (\BdToKstmm)}{\deriv \qsq}\right)^{-1}}
\end{align}
and $A_\text{FB}\equiv \frac{3}{4}S_{6s}$
\cite{Altmannshofer_2009}, can be derived.

\section{Detector and simulation}
\label{sec:experiment}
The \lhcb detector~\cite{LHCb-DP-2008-001,LHCb-DP-2014-002} is a single-arm forward
spectrometer covering the pseudorapidity range $2<\eta <5$, designed for the study of particles containing \bquark or \cquark quarks. The detector used to collect the data analysed in this paper includes a high-precision tracking system consisting of a silicon-strip vertex detector surrounding the \proton\proton interaction region, a large-area silicon-strip detector located upstream of a dipole magnet with a bending power of about $4{\mathrm{\,T\,m}}$, and three stations of silicon-strip detectors and straw drift tubes placed downstream of the magnet.

The tracking system provides a measurement of the momentum, \ptot, of charged particles with
a relative uncertainty that varies from 0.5\% at low momentum to 1.0\% at 200\gevc. The minimum distance of a track to a primary $pp$ collision vertex (PV), the impact parameter (IP), is measured with a resolution of $(15+29/\pt)\mum$, where \pt is the component of the momentum transverse to the beam, in\,\gevc. Different types of charged hadrons are distinguished using information from two ring-imaging Cherenkov detectors. Photons, electrons and hadrons are identified by a calorimeter system consisting of scintillating-pad and preshower detectors, an electromagnetic and a hadronic calorimeter. Muons are identified by a system composed of alternating layers of iron and multiwire proportional chambers. The online event selection is performed by a trigger, which consists of a hardware stage, based on information from the calorimeter and muon systems, followed by a software stage, which applies a full event reconstruction~\cite{LHCb-DP-2012-004}.

Signal candidates are first required to pass the hardware trigger, which selects events containing at least one muon with high \pt. In the subsequent software stages, the decay products are required to have minimum \pt requirements (depending on the run period), in addition to $\text{IP}>100\mum$ with respect to all PVs in the event. This rejects tracks which are in the vicinity of the PV. The final stage of the trigger is based on the topology on the \BdToKstmm decay, featuring at least two tracks that form a good quality vertex significantly displaced from any PV.

Simulated samples are required to model the effects of the detector acceptance and the imposed selection. Collisions of protons are generated using
\pythia~\cite{Sjostrand:2007gs} with an \lhcb configuration~\cite{LHCb-PROC-2010-056}. Decays of unstable particles
are described by \evtgen~\cite{Lange:2001uf}, in which final-state radiation is generated using \photos~\cite{davidson2015photos}.
The interaction of the generated particles with the detector, and its response, are implemented using the \geant
toolkit~\cite{Allison:2006ve, Agostinelli:2002hh} as described in
Ref.~\cite{LHCb-PROC-2011-006}.

Triggered data further undergo a centralised, offline processing step
to deliver analysis-ready datasets across the entire \lhcb physics programme~\cite{Stripping}. Here very loose particle identification (PID) requirements and selections on topological and kinematic variables are applied.

To determine the selection and detector acceptance, simulated samples are generated uniformly in the phase space and processed through the full selection chain. These samples are corrected to account for known data-simulation differences in the \Bd production kinematics and detector occupancy. These corrections are derived from a large sample of \decay{\Bp}{\jpsi\Kp} decays where the background is subtracted using the \sPlot technique~\cite{Pivk:2004ty}. Additional corrections are made to account for track reconstruction, particle identification and trigger inefficiencies, all derived from standardised calibration samples~\cite{LHCb-PUB-2016-021}.

\section{Selection}
\label{sec:Selection}

Signal candidates are required to pass an initial loose selection, which is the first stage of the offline processing. Tracks are required to be well separated and the \Bd decay vertex is required to be of good quality. Furthermore, stringent requirements are applied to the particle identification requirements of the decay products.

Subsequently, dedicated vetoes are applied to remove specific background decays which could otherwise mimic the signal. These include decays where a hadron is misidentified, such as \decay{\Bs}{\phi(1020) (\to\Kp\Km)\mup\mun} and \decay{\Lb}{\proton\Km\mup\mun}. In addition, processes where two hadrons are misidentified, such as \BdToKstmm, where a kaon is misidentified as a pion and vice versa, or \decay{\Lb}{\proton\Km\mup\mu}, where the proton and the kaon are misidentified as a kaon and a pion, respectively, are considered. Decays from \decay{\Bd}{\Kstarz \jpsi} and \decay{\Bd}{\Kstarz \psitwos} where a muon is misidentified as a kaon (pion) and a kaon (pion) is misidentified as a muon are also considered. These peaking backgrounds are suppressed with boosted decision tree~(BDT) classifiers~\cite{Breiman} implemented in \textsc{XGBoost}~\cite{XGBoosting}, using a cross-validation scheme.
Simulated \BdToKstmm decays and simulated samples of each background process are used as the signal and background proxy, respectively. In each BDT the features used are the invariant masses calculated under alternative mass hypotheses for the final state particles and PID information.
The optimal BDT selection criteria are chosen to ensure greater than 99\% signal efficiency. An additional requirement is applied to remove backgrounds from \decay{\Bp}{\Kp\mup\mun} decays with a random pion associated. Candidates in both the mass windows $m_{\Kp\pim\mup\mun}>5380$~\mevcc and $5220<m_{\Kp\mup\mun}<5340$~\mevcc are excluded.

A further selection is performed by means of a BDT classifier to suppress combinatorial background. A signal proxy is taken from \BdToKstmm simulation and a background proxy is taken from the data in the region ${5350 < m_{\Kp\pim\mup\mun} < 5700}$~\mevcc, after the removal of the peaking backgrounds and applying the previous selection requirements. A cross-validation scheme is used to reduce training bias. The features used in the BDT include some PID variables, and kinematic and topological quantities of the intermediate and final state particles. These include the significance of the impact parameter with respect to the primary vertex of the hadron tracks, which was found to be powerful in removing backgrounds from semileptonic decays, e.g. \decay{\Bu}{\Dzb (\to \Kp \mun \neumb) \mup \neum}, as well as reducing the overall level of combinatorial background. The working point of this BDT is determined by maximising the approximate significance $\frac{S}{\sqrt{S+B}}$ where $S$ is the estimated signal yield and $B$ is the estimated combinatorial background yield. The estimated signal yield is obtained from simulation and the background yield from data, by extrapolating the background proxy into the full \Bd mass region.

Two more BDTs are trained to specifically remove \decay{\Bp}{\Kstarp\mup\mun} backgrounds, where neutral particles from \decay{\Kstarp}{\Kp\piz} or \decay{\Kstarp}{\KS\pip} decays are not reconstructed and a random pion or kaon from the primary vertex is selected. These background components exhibit correlations between angular and $m_{\Kp\pim\mup\mun}$ variables. In this case the signal proxy is \BdToKstmm simulation and the background proxies are taken from respective simulation for each background. These BDTs are trained on variables which are related to the reconstruction quality and rapidity of each track, and the quality of the vertices formed. Similarly to the combinatorial BDT, the working points of these BDTs are chosen by maximising the approximate signal significance where the estimated signal yield and yields of the respective backgrounds are used, obtained via simulation.

Candidates are also required to have $5175<m_{\Kp\pim\mup\mun}<5700$~\mevcc, ${796<m_{\kaon\pion}<996}$~\mevcc, and as discussed in Sec.~\ref{sec:angular}, \qsqregion.

\section{Acceptance}
\label{sec:acceptance}

Detector effects, selection, and reconstruction affect the angular and \qsq distributions. These effects are accounted for with a four-dimensional function, which is modelled as
\begin{align}
    \varepsilon(\ctl, \ctk, \phi, \qsq) &= \sum_{ijmn} c_{ijmn} L_i(\ctl) L_j(\ctk) L_m(\phi) L_n(\qsq)
    \label{eqn:acc}
\end{align}
where $L_i$ are Legendre polynomials of order $i$ and $c$ are coefficients used to parameterise the acceptance. The variables $\phi$ and \qsq are mapped to the interval $[-1,1]$.
This acceptance function is obtained from simulation, with events generated uniformly over the phase space, to which the selection is applied. Weights are applied to each data-taking period to correct for data-simulation differences and for the corresponding integrated luminosities. The coefficients $c$ are obtained via the method of moments.
 
\begin{figure}[tb]
  \centering
\includegraphics[width=0.49\linewidth]{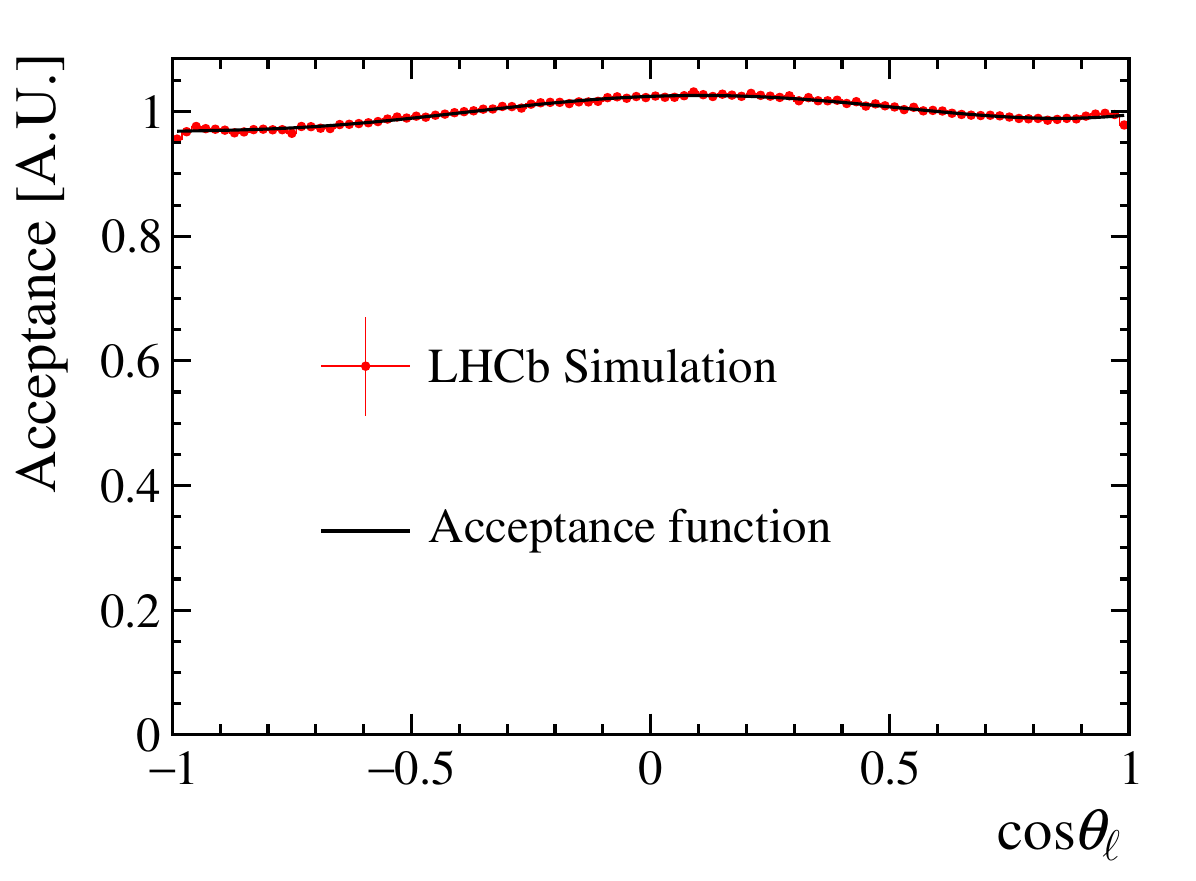}
\includegraphics[width=0.49\linewidth]{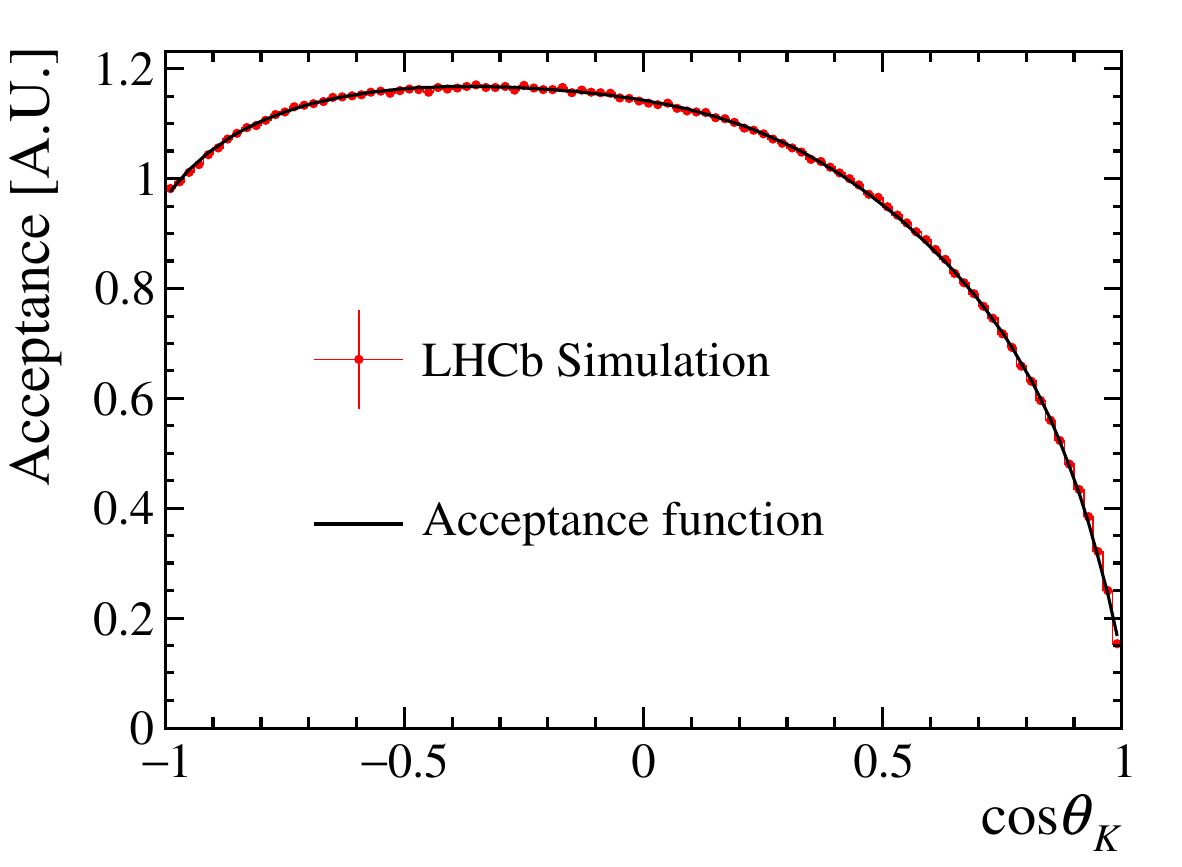}
\includegraphics[width=0.49\linewidth]{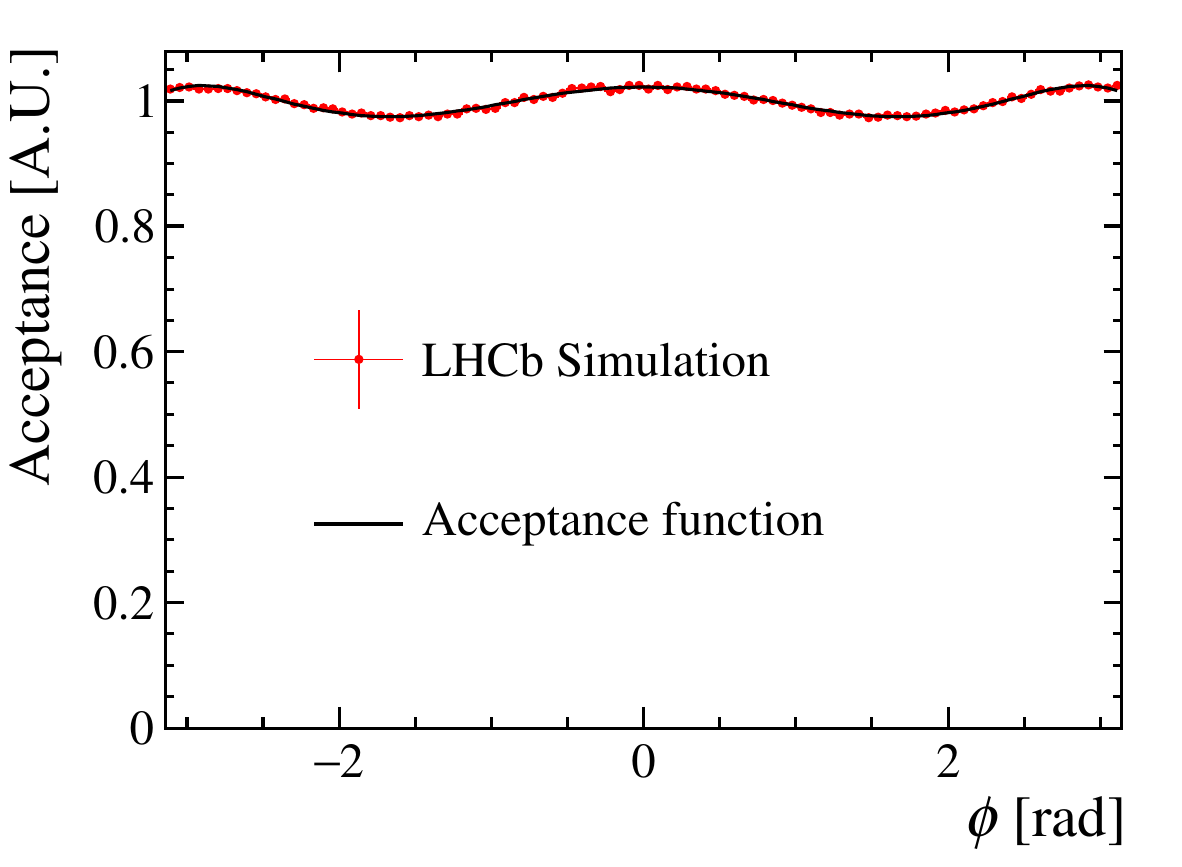}
\includegraphics[width=0.49\linewidth]{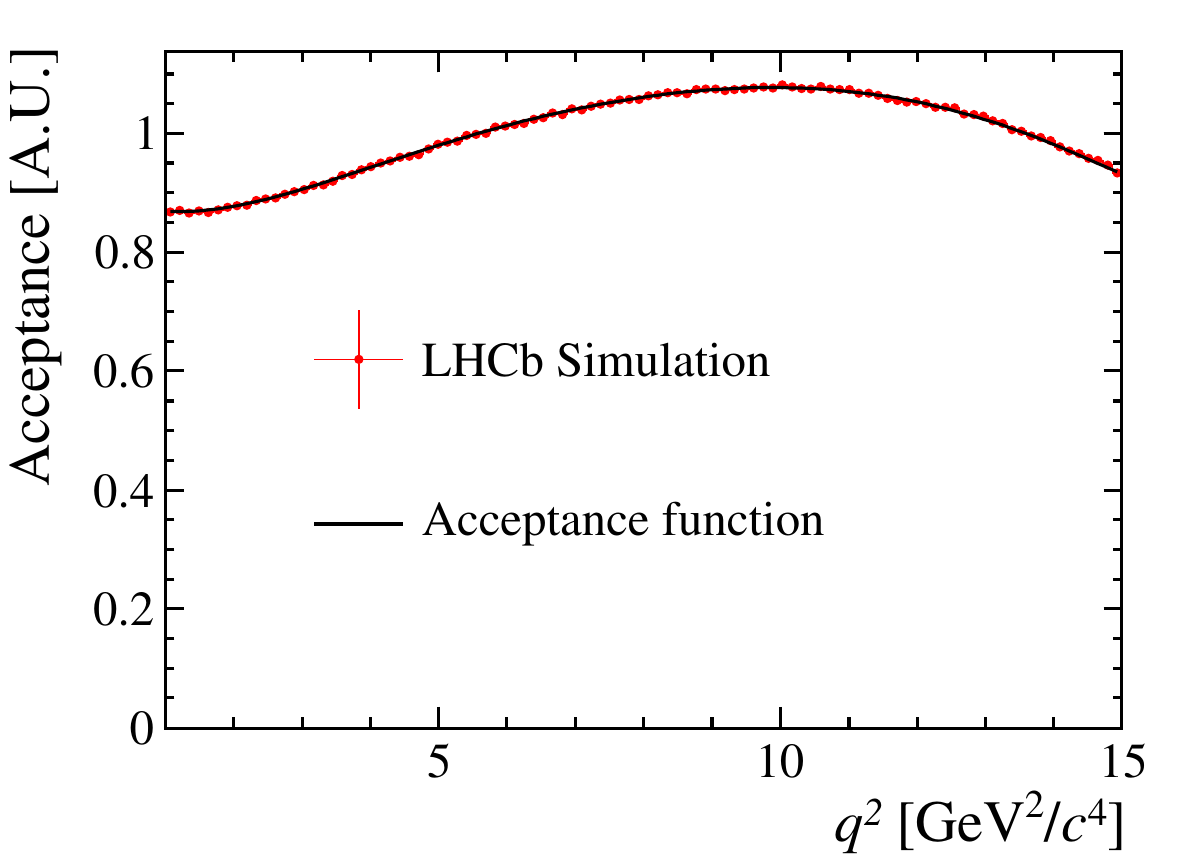}
\caption{
  Acceptance function projected along the four dimensions, (top left)
  $\cos{\theta_\ell}$, (top right) $\cos{\theta_K}$, (bottom left) $\phi$, and (bottom right) $q^2$. The simulated data used to derive the acceptance function is also shown.
}
  \label{fig:acc}
\end{figure}

The maximum number of polynomial orders used for each variable, $[i,j,m,n]_{\textrm{max}}$, is obtained via a machine-learning based goodness-of-fit scheme. For each possible set of maximum polynomial orders, an acceptance function is determined by computing the moments from simulation. Pseudodatasets are then generated from this acceptance function and compared to the simulation. BDTs are used to make this comparison, and are trained to discriminate the simulation from pseudodatasets generated from the acceptance function. These BDTs are constructed via the TMVA~\cite{TMVA4} toolkit, with the three angular variables and \qsq as features. 
For each possible acceptance model, an additional large pseudodataset is generated from the acceptance function, known as the `reference dataset'. BDTs are used to compare the simulation and all other pseudodatasets to this reference dataset. The performance of each classifier is quantified using the area under the receiver operating characteristic (ROC) curve (AUC), which serves as a figure-of-merit. A comparison is made between the distribution of AUCs obtained from the pseudodataset versus reference dataset BDTs and the AUC obtained for the simulation versus reference dataset BDT. From this, a p-value is extracted. The polynomial orders chosen are the lowest for which the p-value exceeds 5\%. This yields an optimal maximum number of orders of 6 (\ctl), 8 (\ctk), 6 ($\phi$), and 6 (\qsq).
Plots of the acceptance, projected onto the four variables are shown in Fig.~\ref{fig:acc}. Shown is the corrected simulation with the acceptance function overlaid. Systematic uncertainties associated to the acceptance correction are discussed in Sec.~\ref{sec:systematics}.

\section{Fit strategy}
\label{sec:fit}

After applying the selection described in Sec.~\ref{sec:Selection}, the only significant contributions expected to remain in the data are from the signal decay and residual combinatorial background.
An unbinned extended maximum-likelihood fit is performed. The probability density function (PDF) of the signal model corresponds to the four-differential decay distribution (Eq.~\ref{eqn:total}) multiplied by the acceptance function (Eq.~\ref{eqn:acc}). Since the signal PDF is directly proportional to bilinear combinations of amplitudes, whose absolute size has no physical meaning, its overall scale is arbitrary and is adjusted to correctly reproduce the observed signal yield.

In order to separate the signal component from the remaining background component, the mass of the decay products $m_{\Kp\pim\mup\mun}$ is used. The signal PDF in this variable is parameterised with the sum of two Crystal Ball functions~\cite{Skwarnicki:1986xj} where the peak positions are fixed to each other. In addition, \decay{\Bs}{\Kstarzb(\to\Km\pip)\mup \mun} decays give a non-negligible contribution. This is accounted for in the \Kpmumu mass fit by the sum of two Crystal Ball functions with identical parameters as the \BdToKstmm lineshape, but where the means are shifted by the known $\Bs-\Bd$ mass difference~\cite{PDG2024}. The relative yield is fixed based on the relative branching fractions~\cite{PDG2024} and differences in the hadronisation fractions~\cite{LHCb-PAPER-2020-046}, corresponding to 0.8\% of the \BdToKstmm yield. The angular distribution of the  \decay{\Bs}{\Kstarzb(\to\Km\pip)\mup \mun} decay is assumed to be the same as the signal angular distribution, with the relevant J terms conjugated according to Ref.~\cite{PhysRevD.93.054008}. This assumption holds in the minimal flavour violation paradigm~\cite{D_Ambrosio_2002}.

The \Kpmumu mass distribution for the combinatorial background is described by an exponential function. Regarding the other variables (\ctl, \ctk, $\phi$, and \qsq), Chebyshev polynomials up to second order are used to describe the background. The background PDFs are assumed to factorise, with no correlations between each other or with $m_{\Kp\pim\mup\mun}$. The region of phase space removed by the mass cut to eliminate the \decay{\Bp}{\Kp\mup\mun} contribution is accounted for when normalising the background PDF. The \Bd mass lineshape and background parameters are treated as nuisance parameters.

The fit procedure is validated both with ensembles of pseudoexperiments and using the \BdToJPsiKst control mode. A goodness-of-fit test is performed on the data, using a similar methodology to that used in Sec.~\ref{sec:acceptance}, and it shows that the fit is of good quality.

The validity and stability of the fit is assessed using ensembles of pseudoexperiments.
The uncertainties on the amplitude components are extracted by fitting the profile log-likelihood for each parameter of interest and are therefore allowed to be asymmetric. The correlation matrix of the statistical uncertainties at the best fit point is computed from the Hessian matrix of the likelihood at the minimum.
In order to avoid experimenter's bias, the results of the analysis were not examined until the full procedure had been finalised.

\section{Systematic uncertainties}
\label{sec:systematics}

Systematic uncertainties are computed by generating ensembles of pseudoexperiments. For each source of systematic uncertainty an alternative model or an alternative set of corrections is used to generate pseudodata, which are then fitted with the nominal model. The resulting systematic uncertainty is computed by combining the mean of the deviations and the root-mean squared of the deviations in quadrature, with the full covariances between the parameters preserved. For sources affecting the acceptance, alternative acceptance functions are derived. All systematic sources are treated independently, and their covariance matrices are summed to obtain the total systematic covariance matrix.
Systematic uncertainties fall in two categories: those arising from uncertainties in the modelling in the fits and those arising from uncertainties in the acceptance.

Uncertainties related to the models used in the fit to data are computed by generating pseudodata from alternative models. Some parameters in the $m_{\Kp\pim\mup\mun}$ distribution, namely the tail parameters in the Crystal Ball functions, are fixed to those obtained from fits to \decay{\Bd}{\Kstarz\jpsi}data. For the corresponding systematic uncertainty, the parameters are obtained from fits to \BdToKstmm simulation. The P-S interference terms are sensitive to the parameterisation of the S-wave lineshape in $m_{\kaon\pion}$. An uncertainty associated to this is evaluated by considering an alternative model based on the isobar parameterisation~\cite{LHCb-PAPER-2016-012} in contrast to the default LASS parameterisation~\cite{LASS}. 
The effect of the assumption of minimal flavour violation on the angular distribution of \decay{\Bs}{\Kstarzb \mup \mun} is assessed by replacing it with a model generated by the default \texttt{flavio}~\cite{straub:2018flavio} SM predictions.
The assumption that for the background, the $m_{K\pi\mu\mu}$ and the angular variables are uncorrelated is employed in the fit. Its related uncertainty is tested using a dataset where candidates with muons of the same sign ($K\pi\mu^\pm \mu^\pm$) are selected. The associated systematic contribution is evaluated by sampling with replacement this dataset to the expected background yield. 

Data-simulation differences arising from mismodelled \Bd kinematics and hardware trigger efficiency are corrected using \decay{\Bp}{\jpsi\Kp} data and simulation samples. Possible systematic uncertainties are evaluated by comparing the reference corrections with those obtained using the decay \BdToJPsiKst as an alternate control mode. For the track corrections, the weights are varied according to their uncertainties. In addition, a systematic uncertainty is assigned based on the treatment of the occupancy corrections. The uncertainty due to the size of the simulation used to derive the acceptance function is also estimated by bootstrapping the simulation and computing alternative acceptances.

Table~\ref{tab:syst_breakdown_sources} shows the ratio of systematic to statistical uncertainties. The quoted range covers all of the amplitude components.
The ratio between the total systematic uncertainty $\sigma_{\textrm{syst},\,\textrm{tot}}$ and $\sigma_{\textrm{stat}}$ varies in the range $0.09-0.35$.
The full covariances are provided on the HEPData repository \cite{HEPdata}. The largest systematic uncertainties are those associated with the kinematic and trigger corrections.

\begin{table}
\centering
\caption{Ranges of the ratio between the systematic and statistical uncertainties $\sigma_\mathrm{syst}/\sigma_\mathrm{stat}$ for each systematic source, where the range is taken over all amplitude components.}
\label{tab:syst_breakdown_sources}
\begin{tabular}{lc} \hline 
Systematic source &  $\sigma_\mathrm{syst}/\sigma_\mathrm{stat}$ \\
\midrule
            $\Bd$ mass lineshape & $0.001-0.037$ \\
        $m_{\kaon\pion}$ model & $ 0.001-0.078$ \\
        \decay{\Bs}{\Kstarzb \mup \mun} angular model & $ 0.084-0.118$ \\
        Background non-factorisation & $ 0.004-0.187$ \\
        Kinematic corrections & $ 0.003-0.300$ \\
        Trigger corrections & $ 0.001-0.231$ \\
      Tracking corrections & $ 0.000-0.017$ \\ 
      Treatment of occupancy corrections & $ 0.000-0.014$ \\
    Simulation size & $ 0.003-0.033$ \\  \hline
            Total & $ 0.092-0.347$ \\
\bottomrule
\end{tabular}
\end{table}

\section{Results}
\label{sec:results}

\begin{figure}
    \centering
    \includegraphics[width=0.45\linewidth]{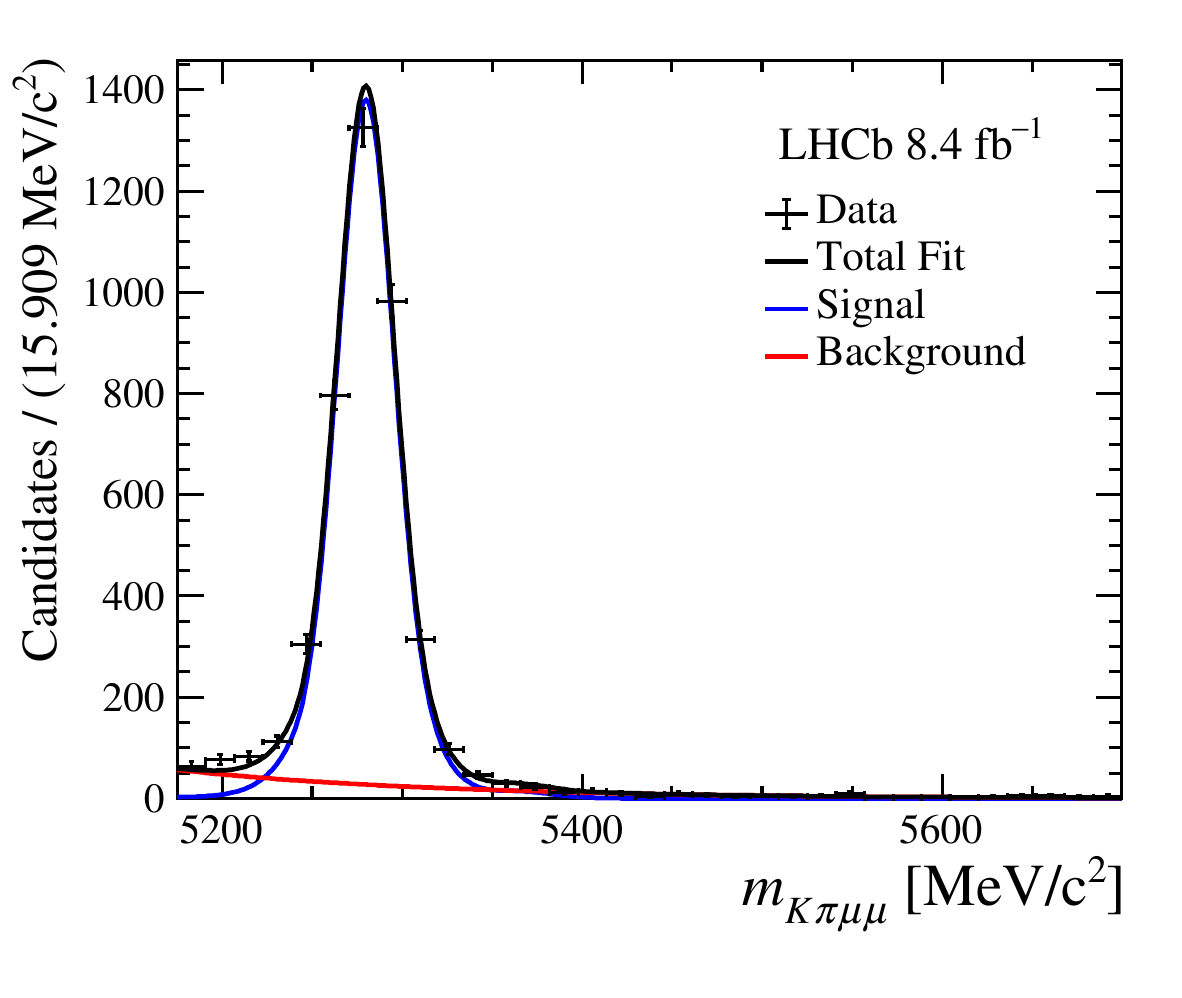}
    \includegraphics[width=0.45\linewidth]{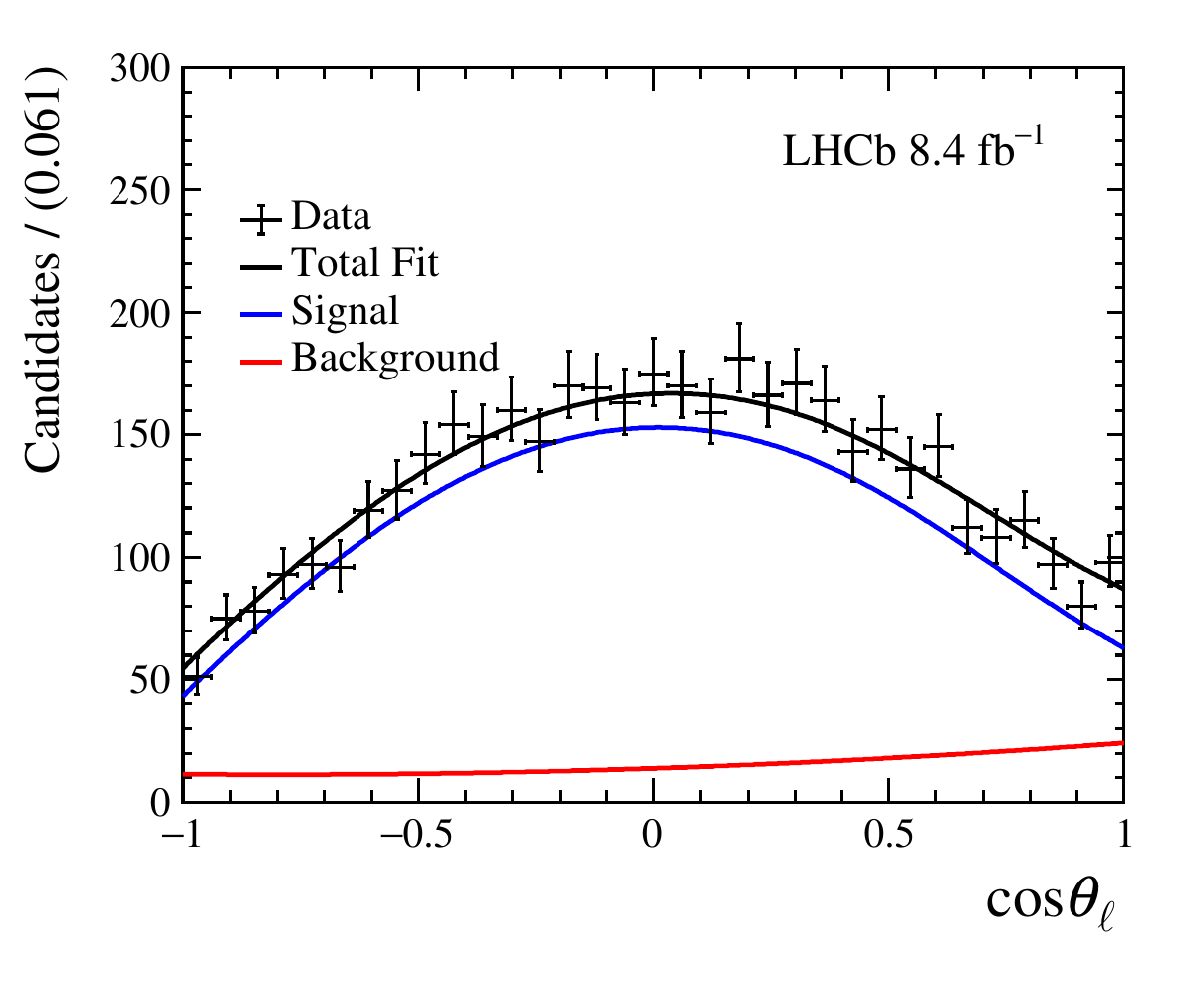}
    \includegraphics[width=0.45\linewidth]{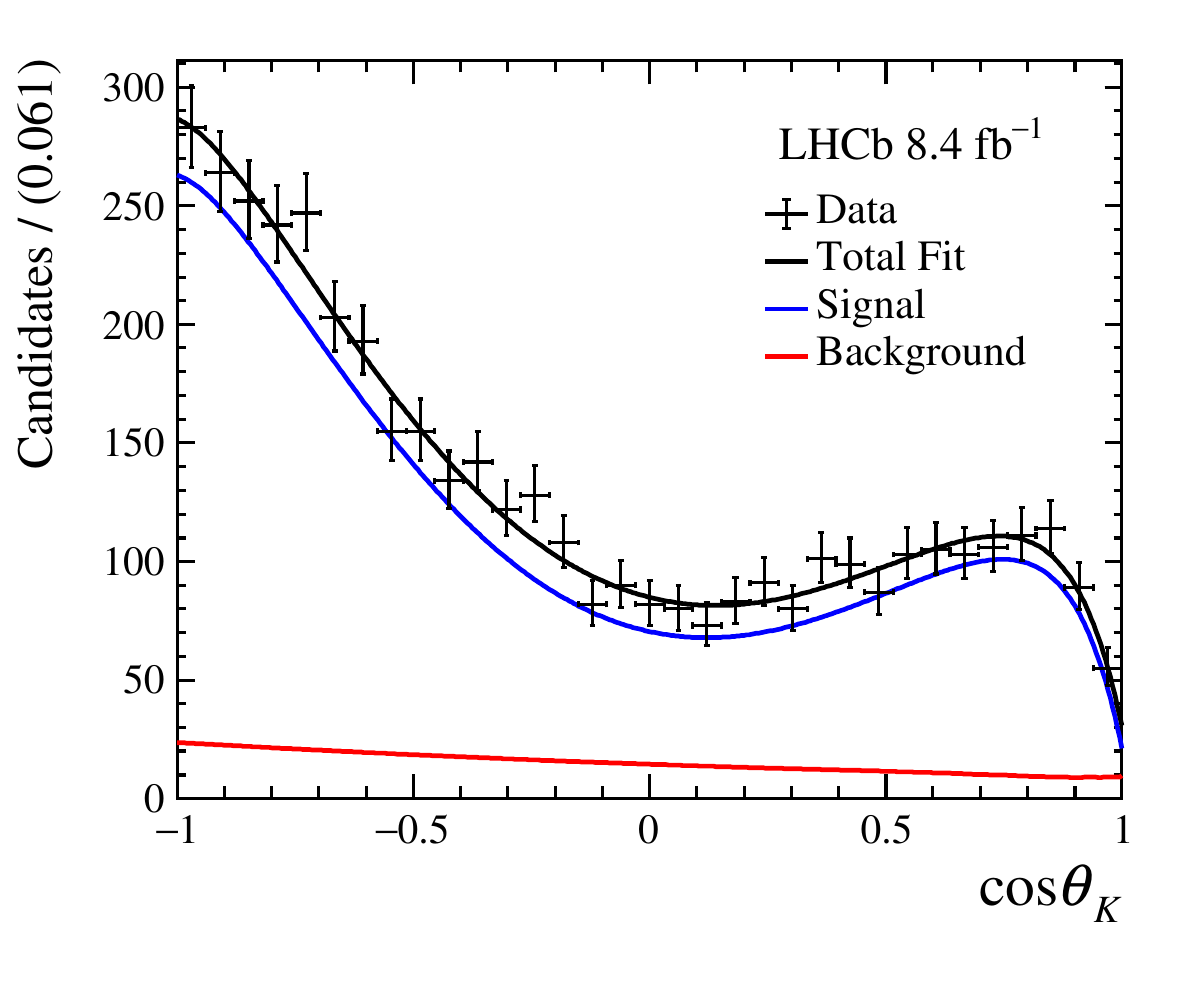}
    \includegraphics[width=0.45\linewidth]{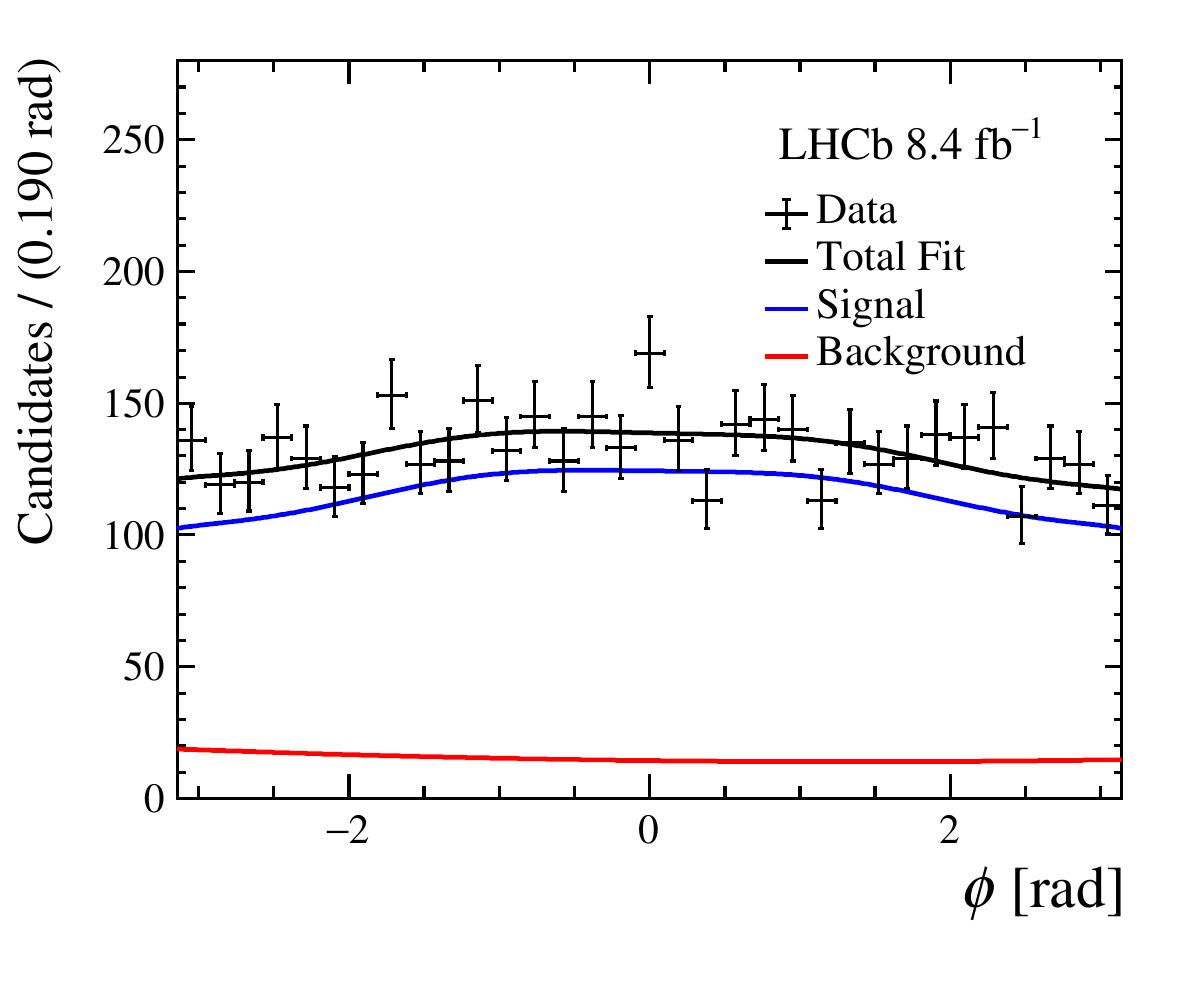}
    \includegraphics[width=0.45\linewidth]{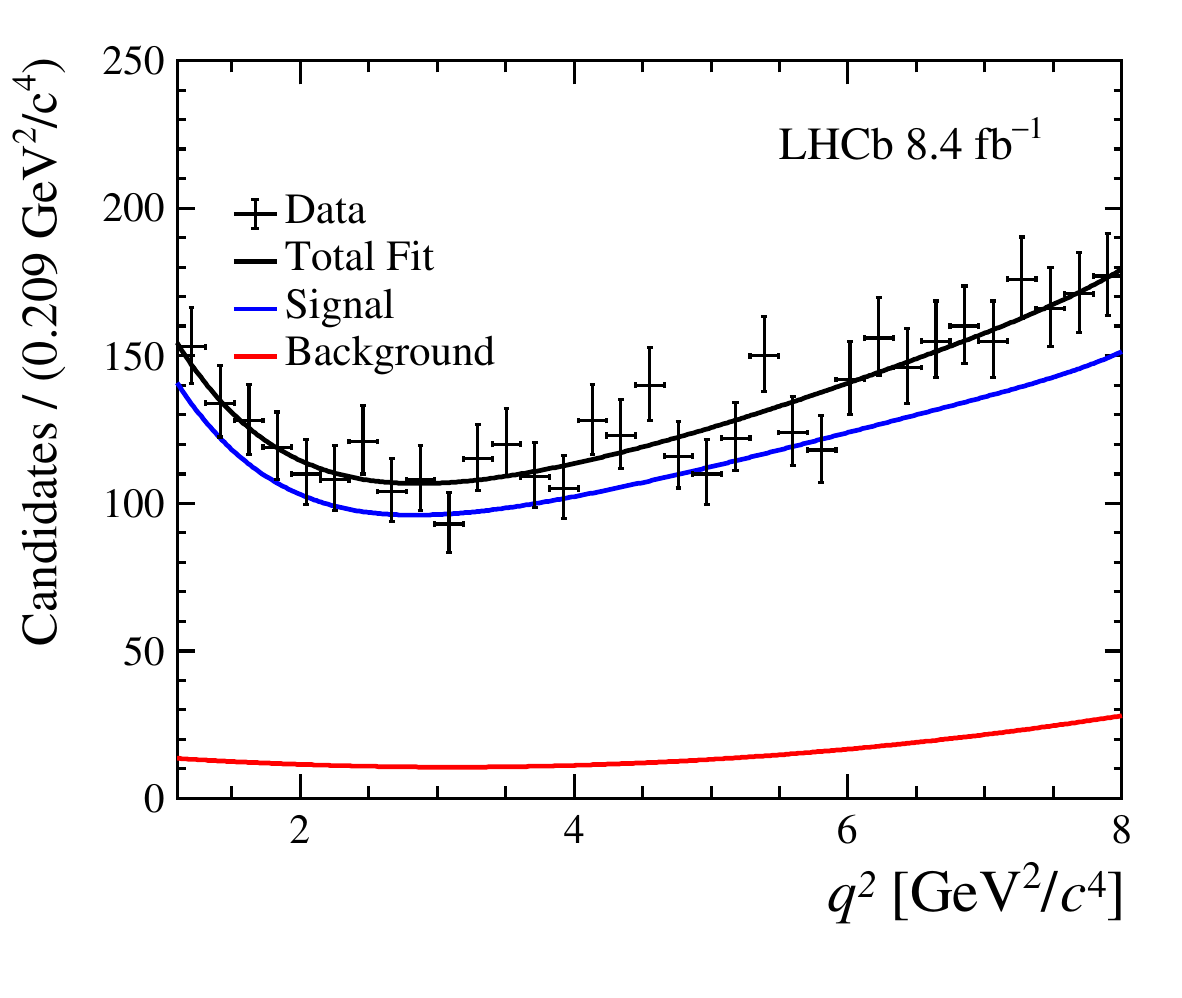}
    \caption{Distributions of the reconstructed (top left) $B^0$ mass, (top right) $\cos{\theta_\ell}$, (middle left) $\cos{\theta_K}$, (middle right) $\phi$, and (bottom) $q^2$. Shown are the data, as well as the fit results with the (blue) signal and (red) combinatorial background components.}
    \label{fig:fitprojections}
\end{figure}

\begin{figure}
    \centering
    \includegraphics[width=0.45\linewidth]{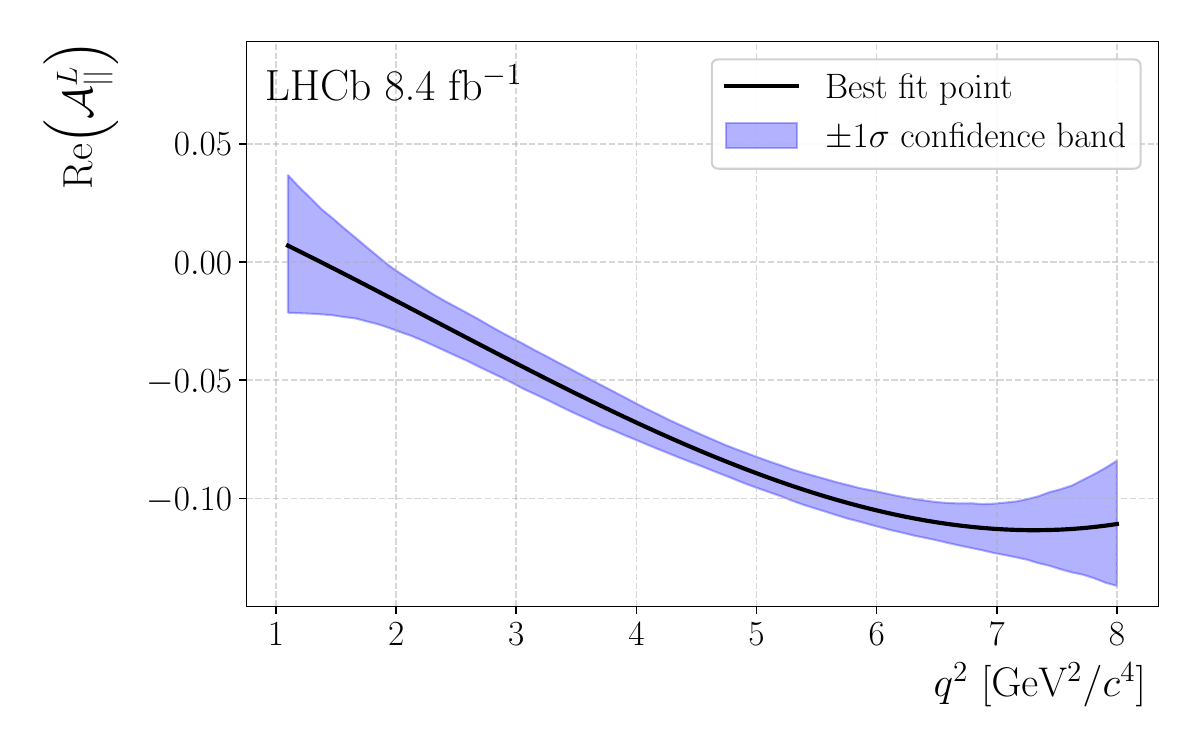}
    \includegraphics[width=0.45\linewidth]{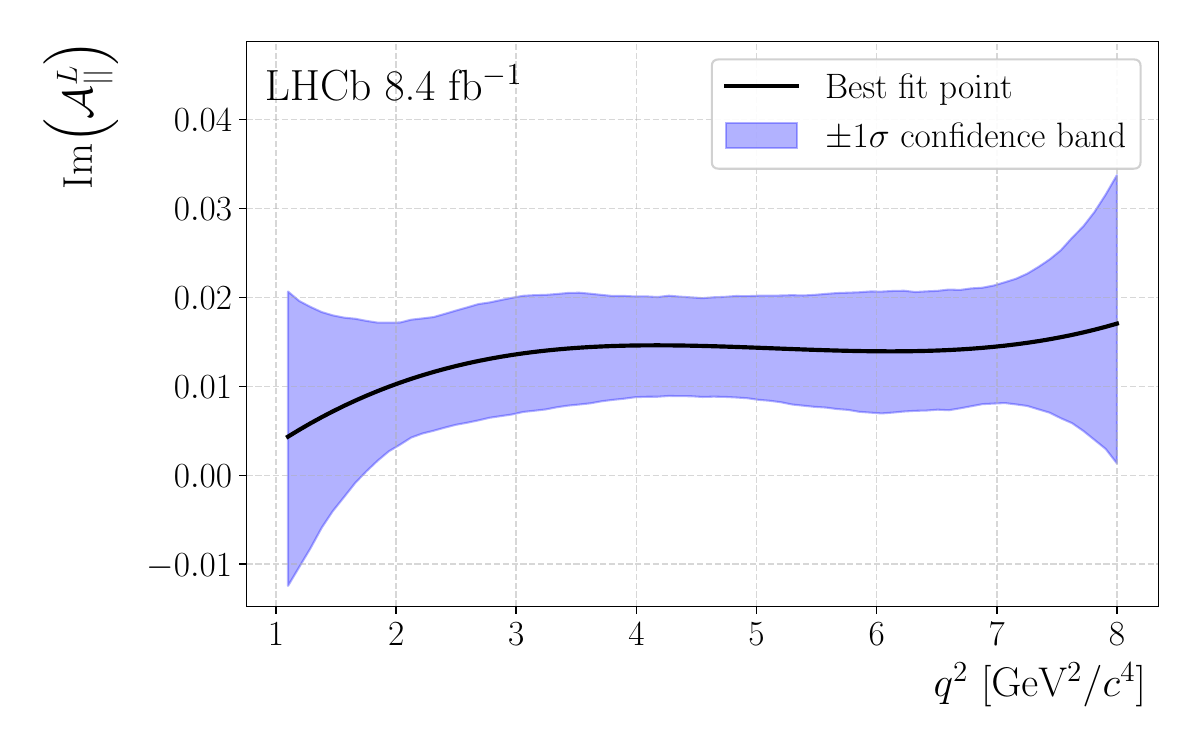}
    \includegraphics[width=0.45\linewidth]{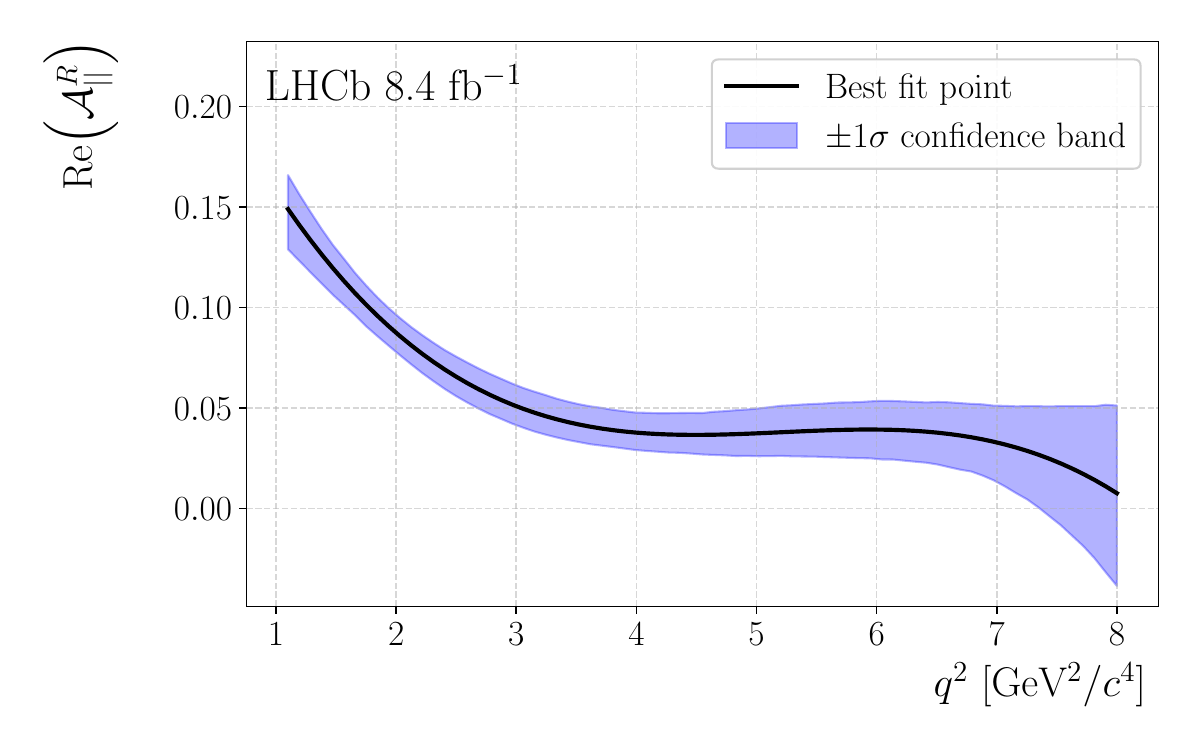}
    \includegraphics[width=0.45\linewidth]{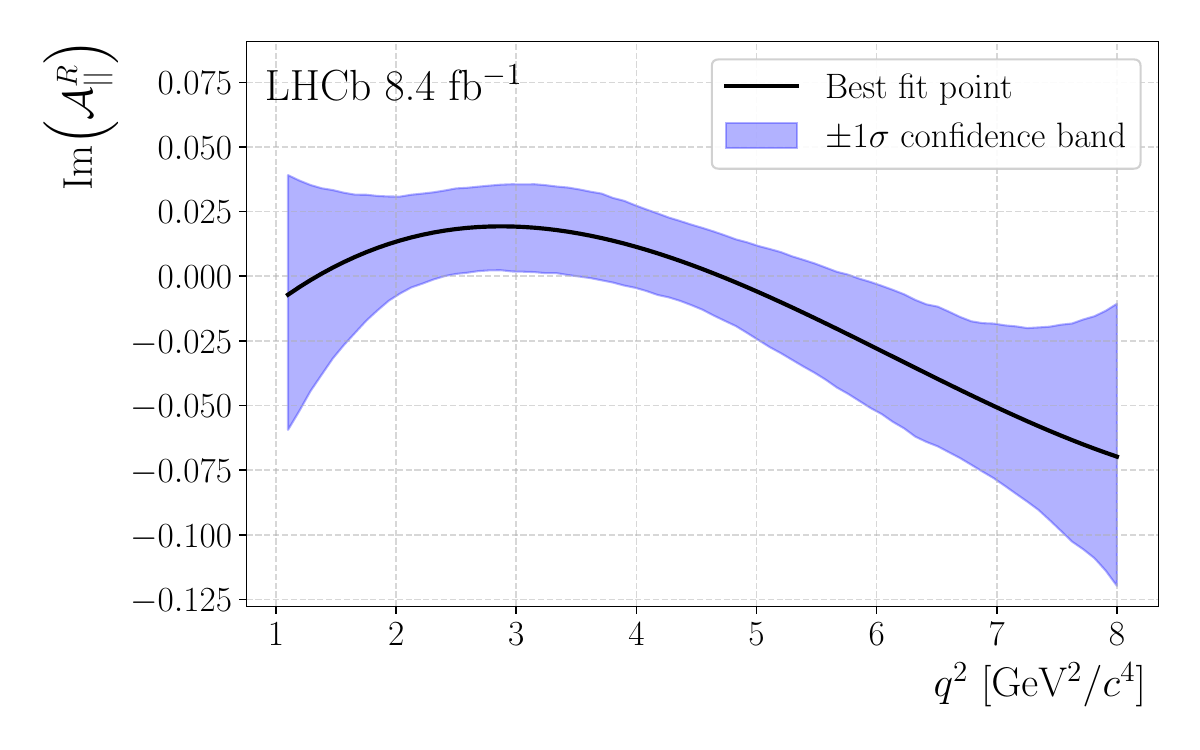}
    \includegraphics[width=0.45\linewidth]{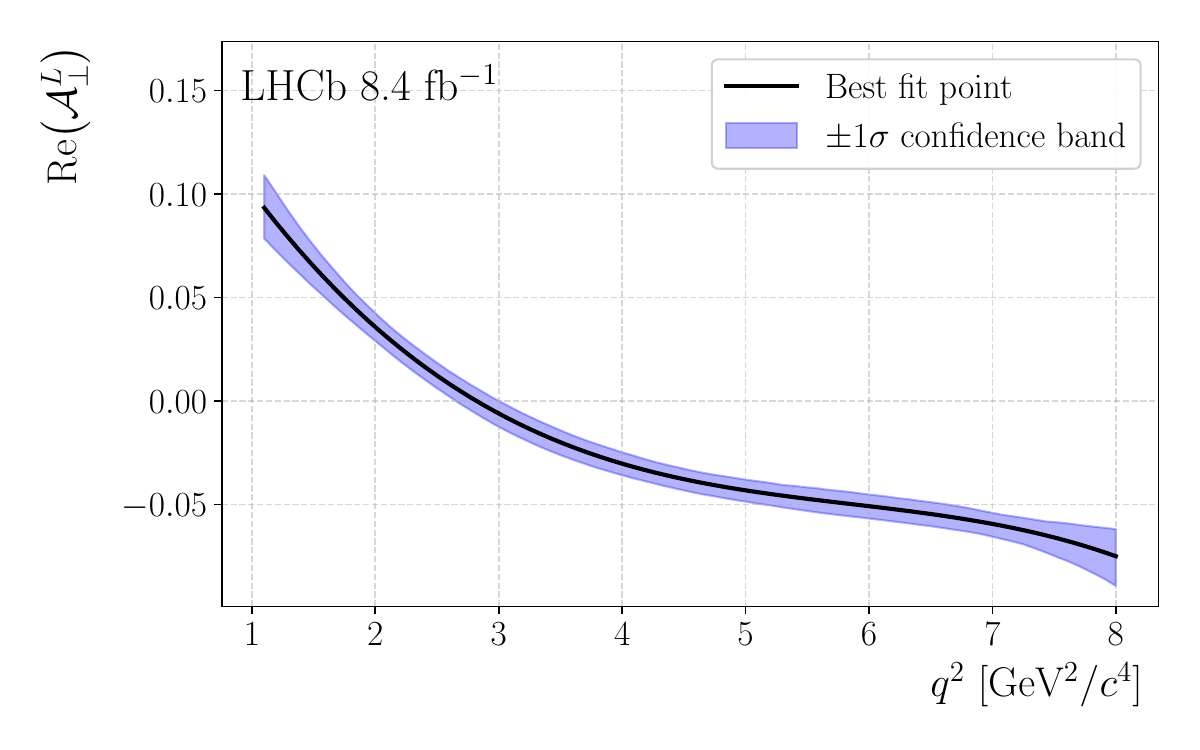}
    \includegraphics[width=0.45\linewidth]{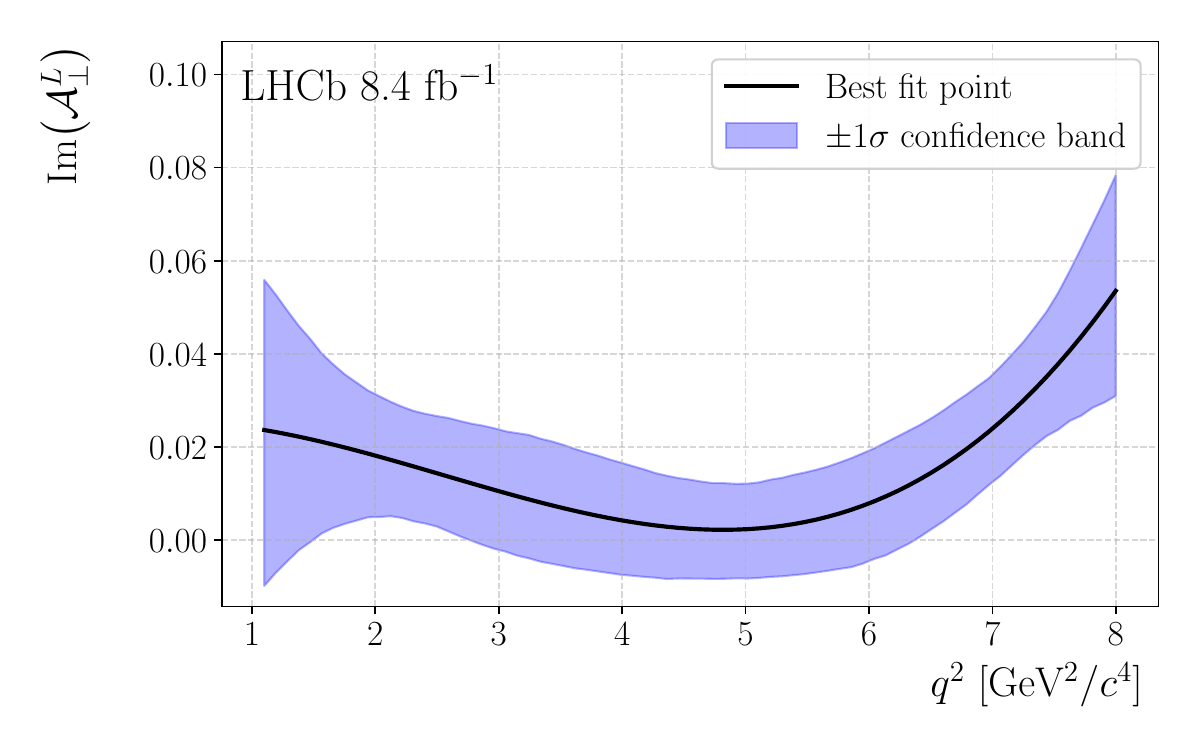}
    \includegraphics[width=0.45\linewidth]{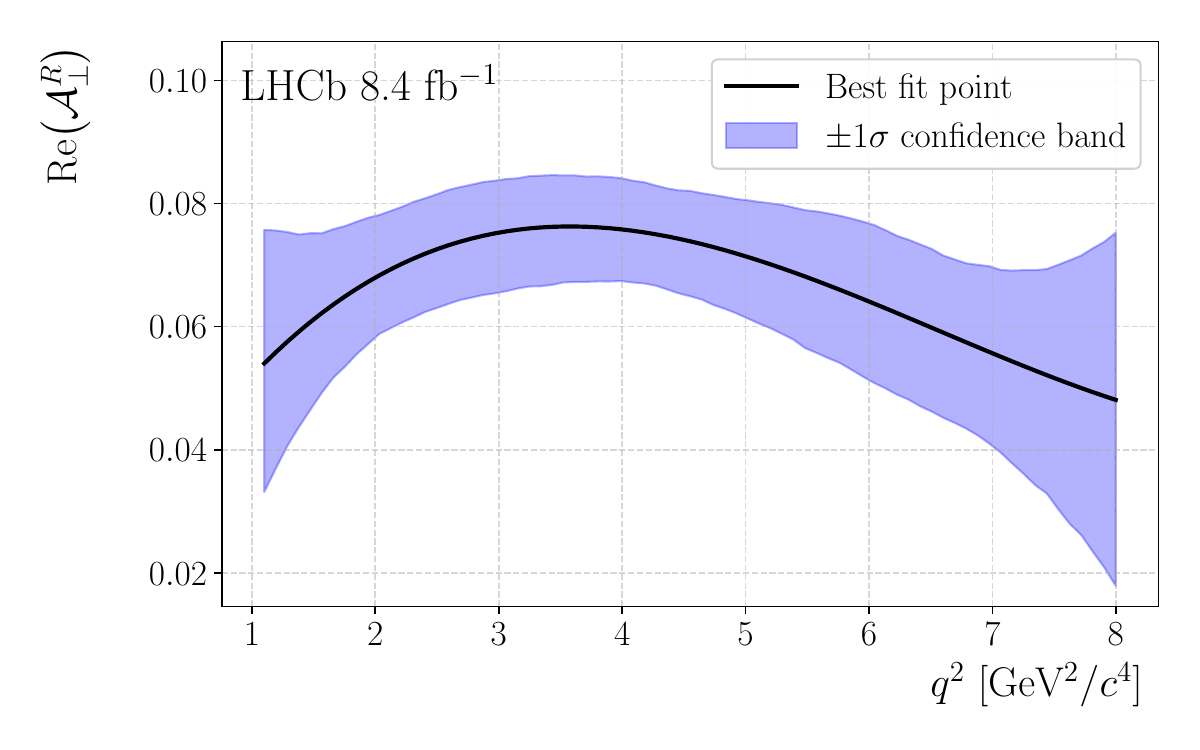}
    \includegraphics[width=0.45\linewidth]{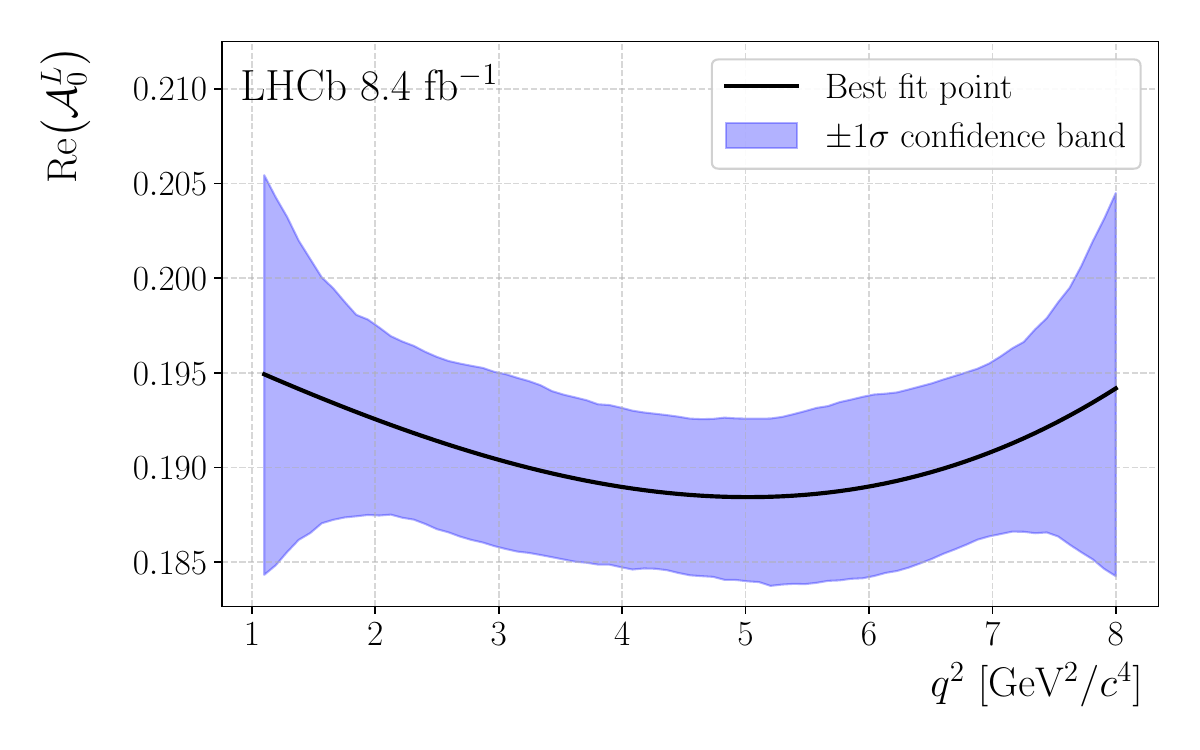}
    \caption{Transversity amplitudes as functions of the $\mumu$ invariant mass squared, $q^2$. Shown are the (black line) best fit points with (blue) $\pm1\sigma$ bands obtained from sampling the total covariance matrix from the fit.}
    \label{fig:fitresults1}
\end{figure}

\begin{figure}
    \centering
    \includegraphics[width=0.45\linewidth]{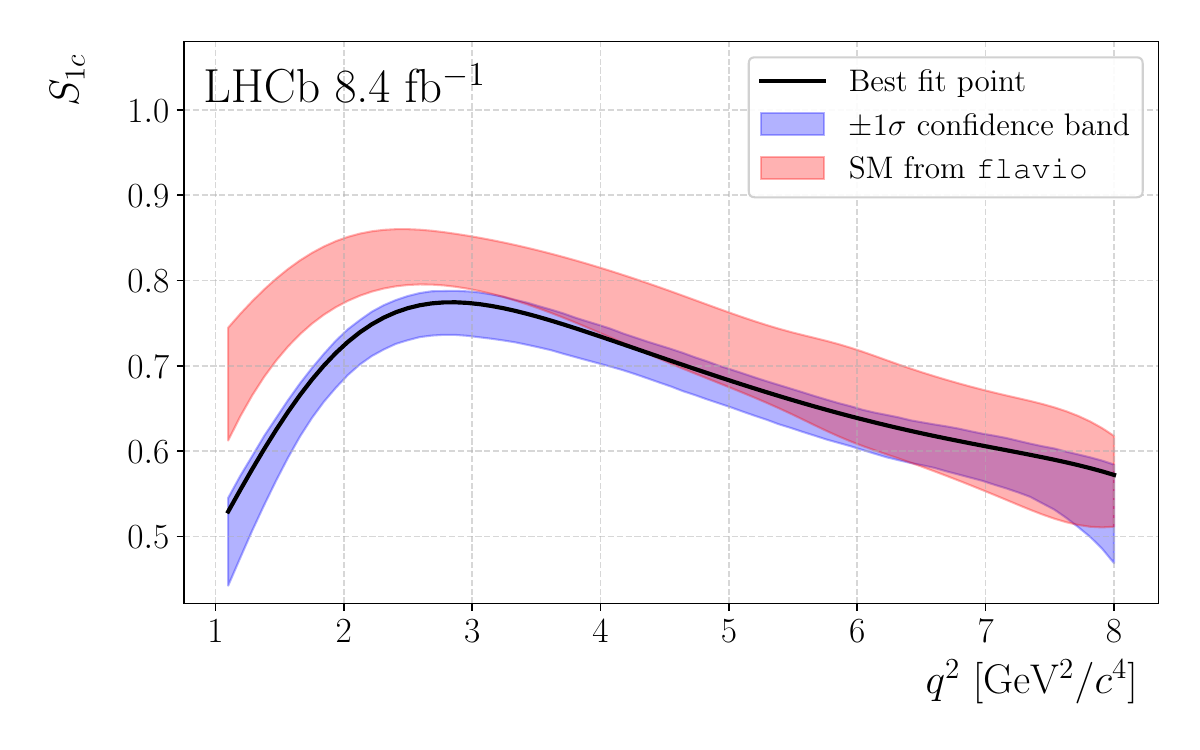}
    \includegraphics[width=0.45\linewidth]{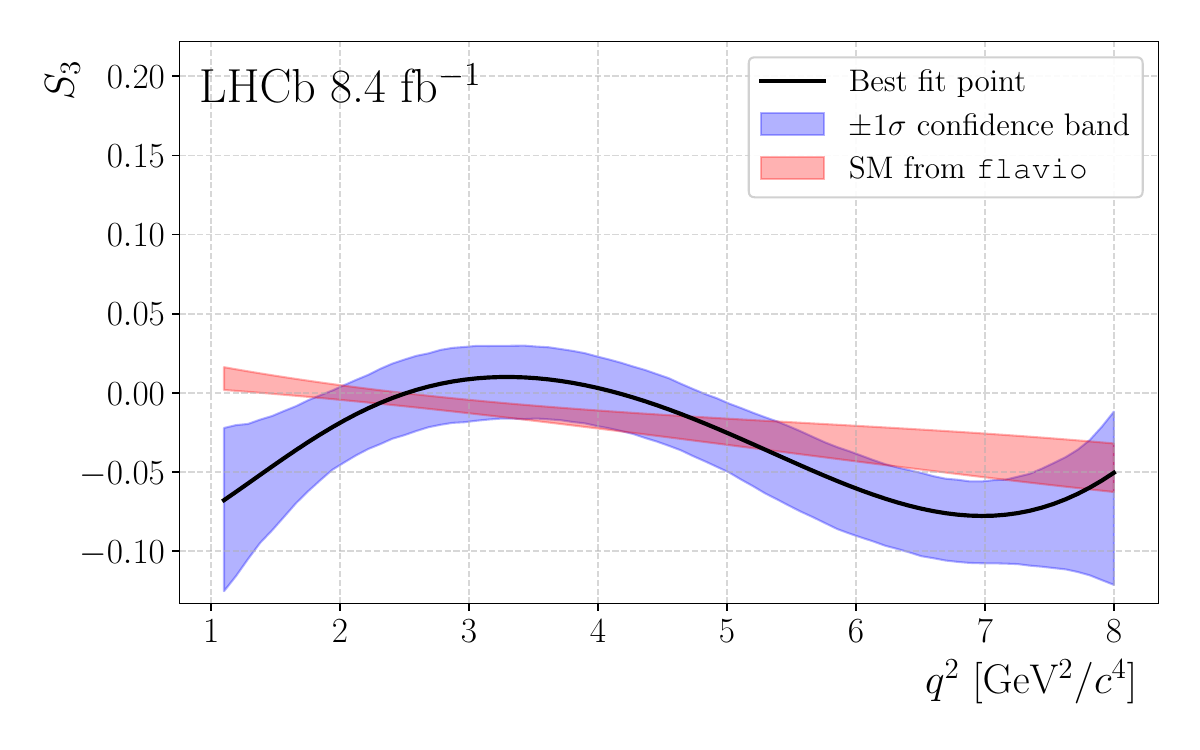}
    \includegraphics[width=0.45\linewidth]{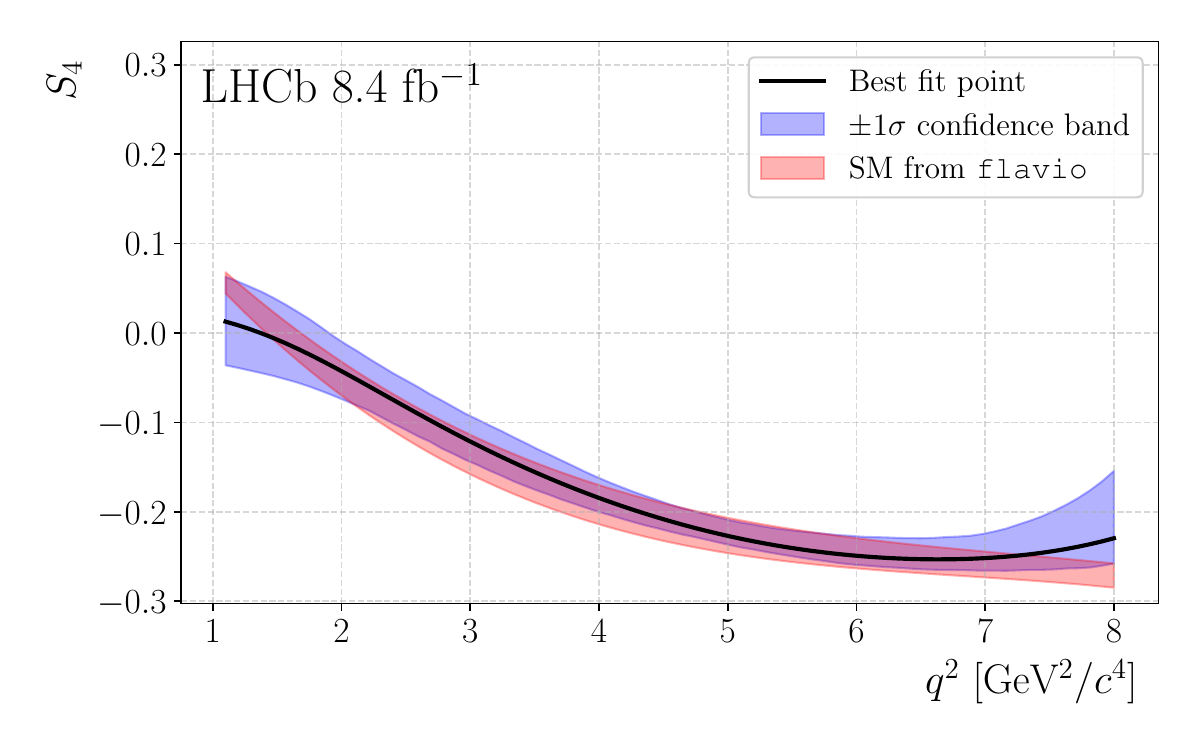}
    \includegraphics[width=0.45\linewidth]{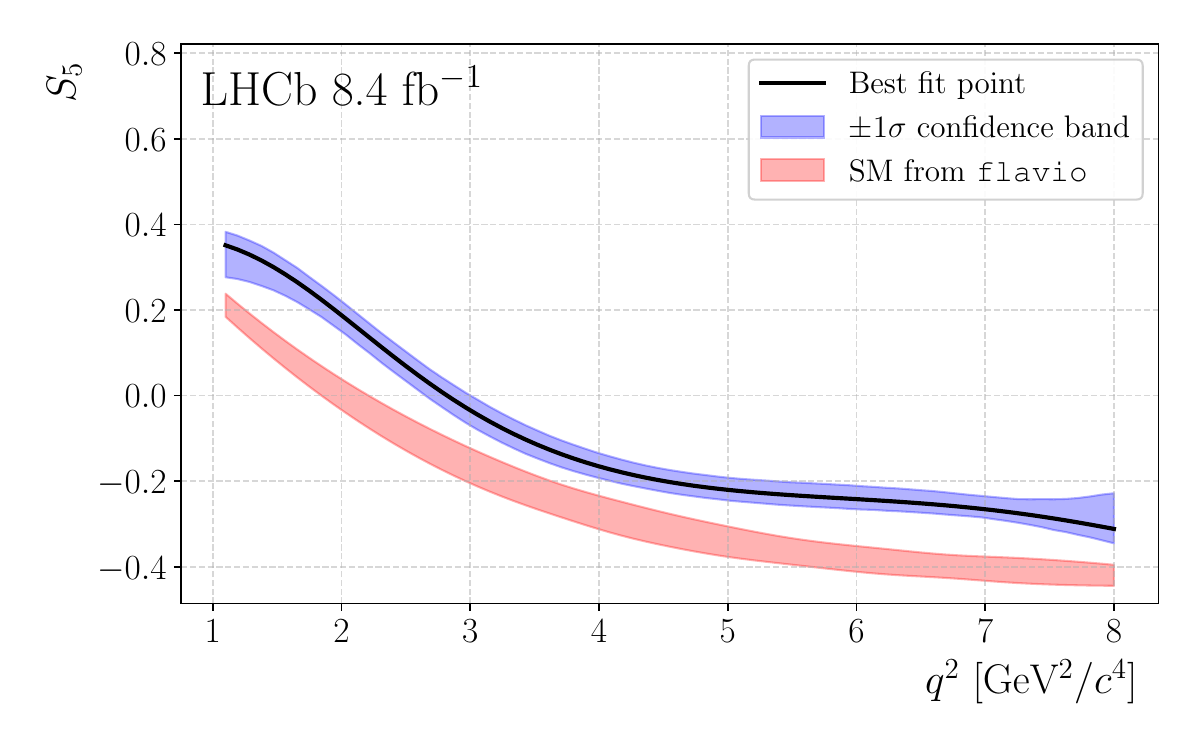}
    \includegraphics[width=0.45\linewidth]{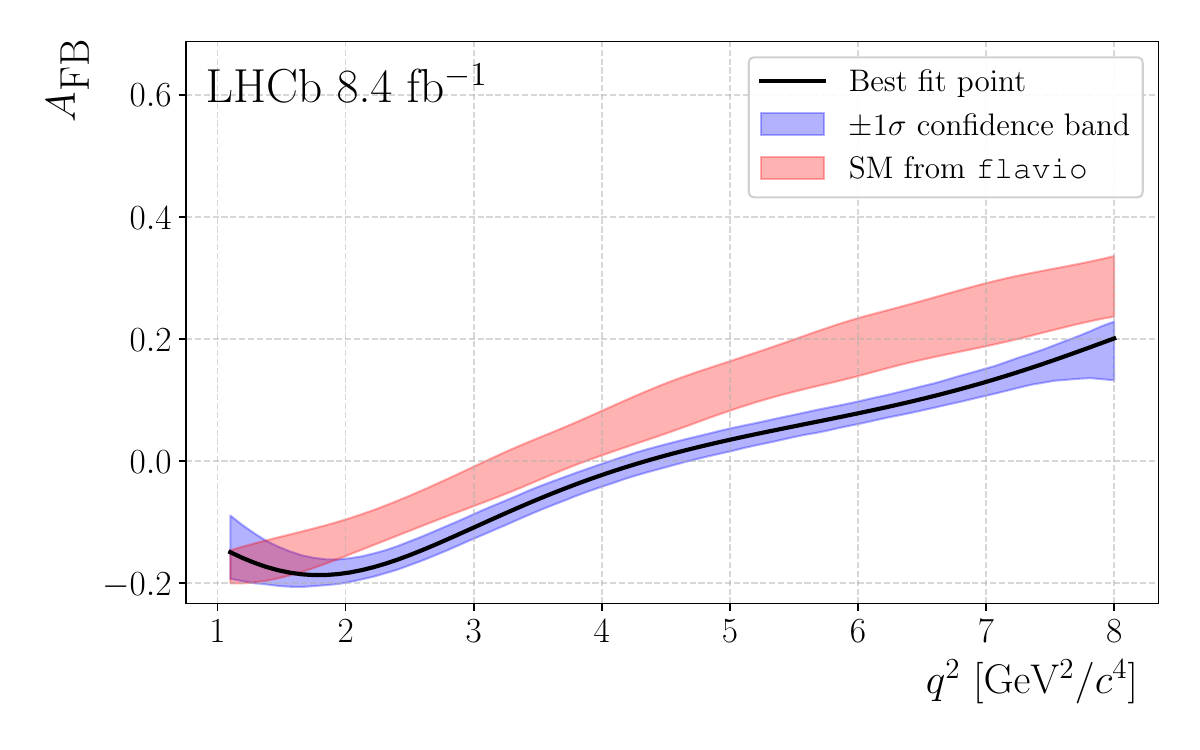}
    \includegraphics[width=0.45\linewidth]{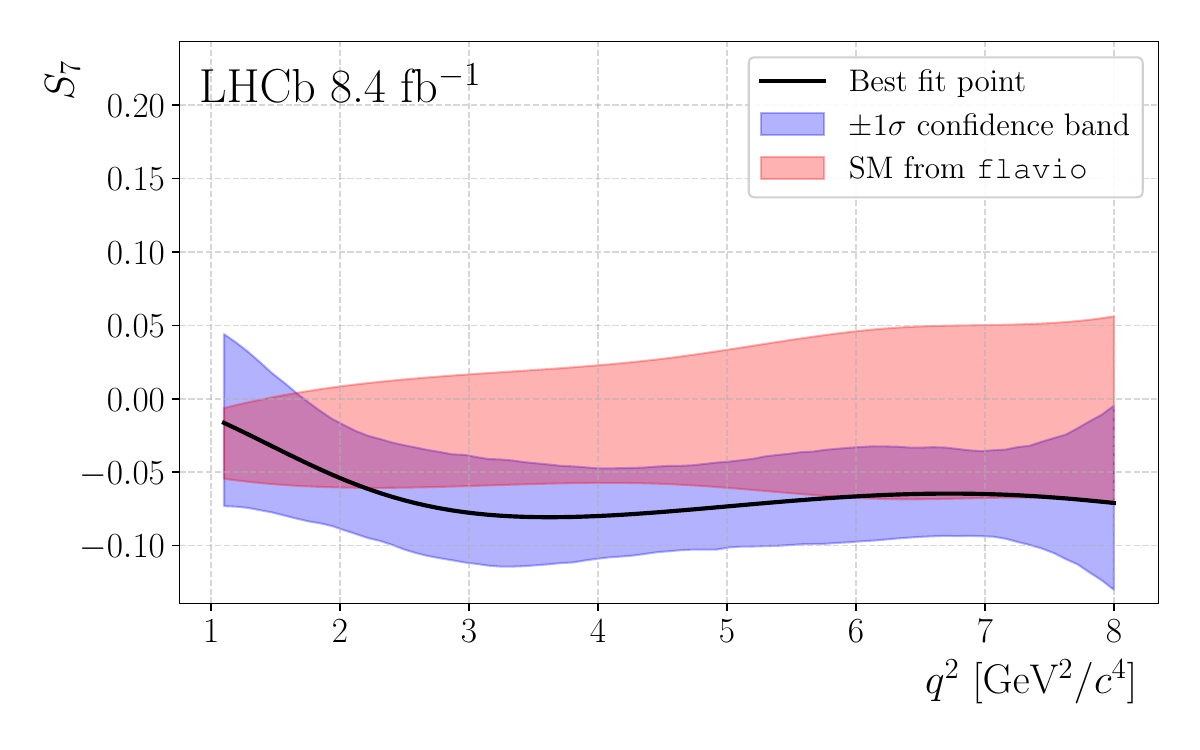}
    \includegraphics[width=0.45\linewidth]{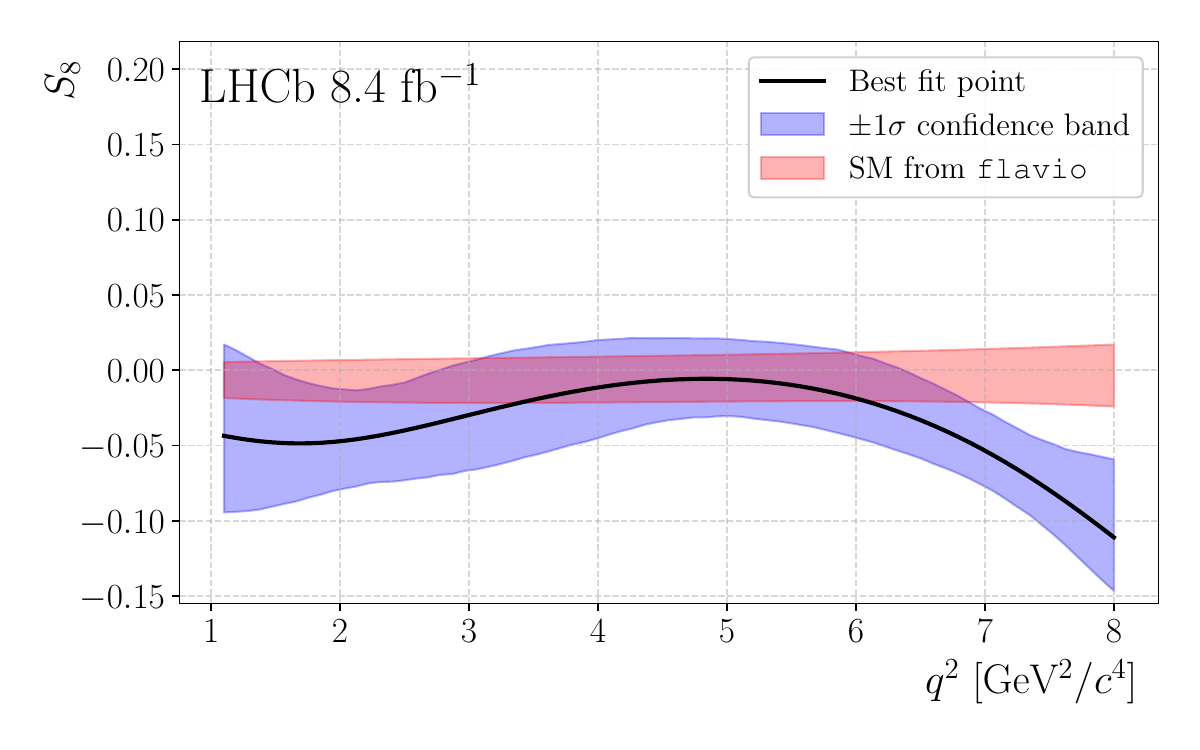}
    \includegraphics[width=0.45\linewidth]{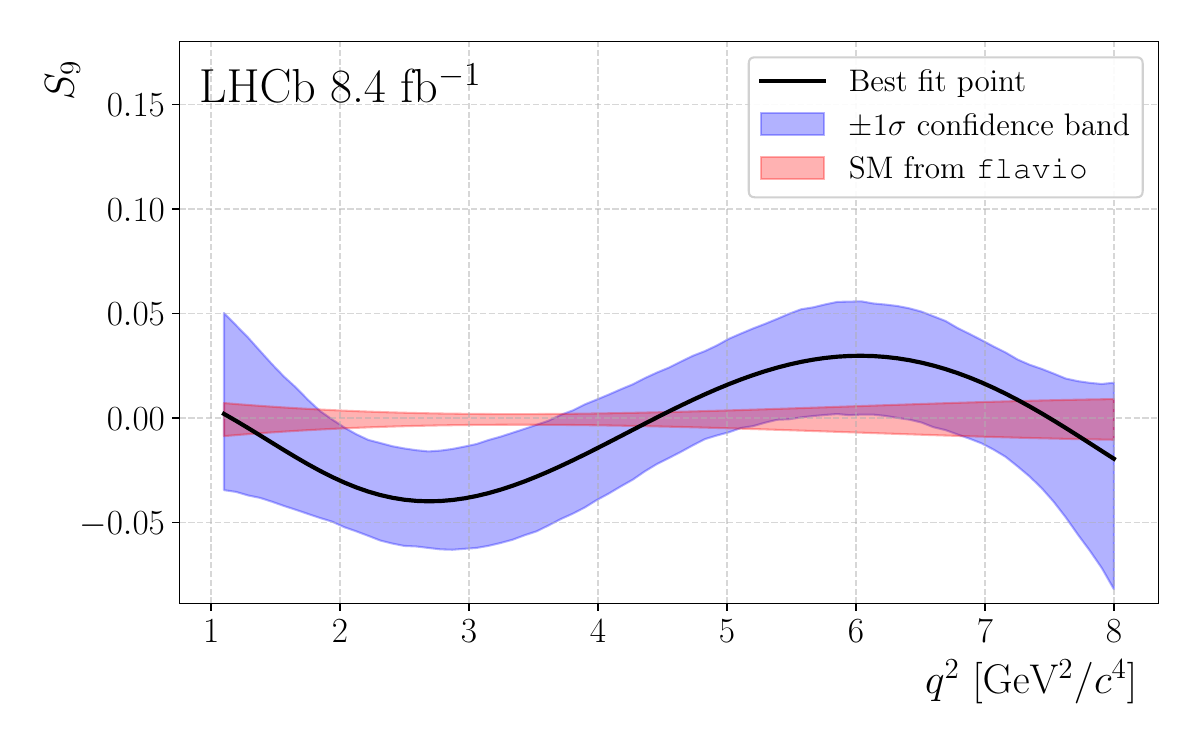}
    \caption{Angular observables as functions of $\mumu$ invariant mass squared, $q^2$. Shown are the (black line) best fit points with (blue) $\pm1\sigma$ bands obtained from sampling the total covariance matrix from the fit. The predictions obtained from \texttt{flavio} with SM parameters are also shown.}
    \label{fig:fitresults2}
\end{figure}

\begin{figure}
    \centering
    \includegraphics[width=0.49\linewidth]{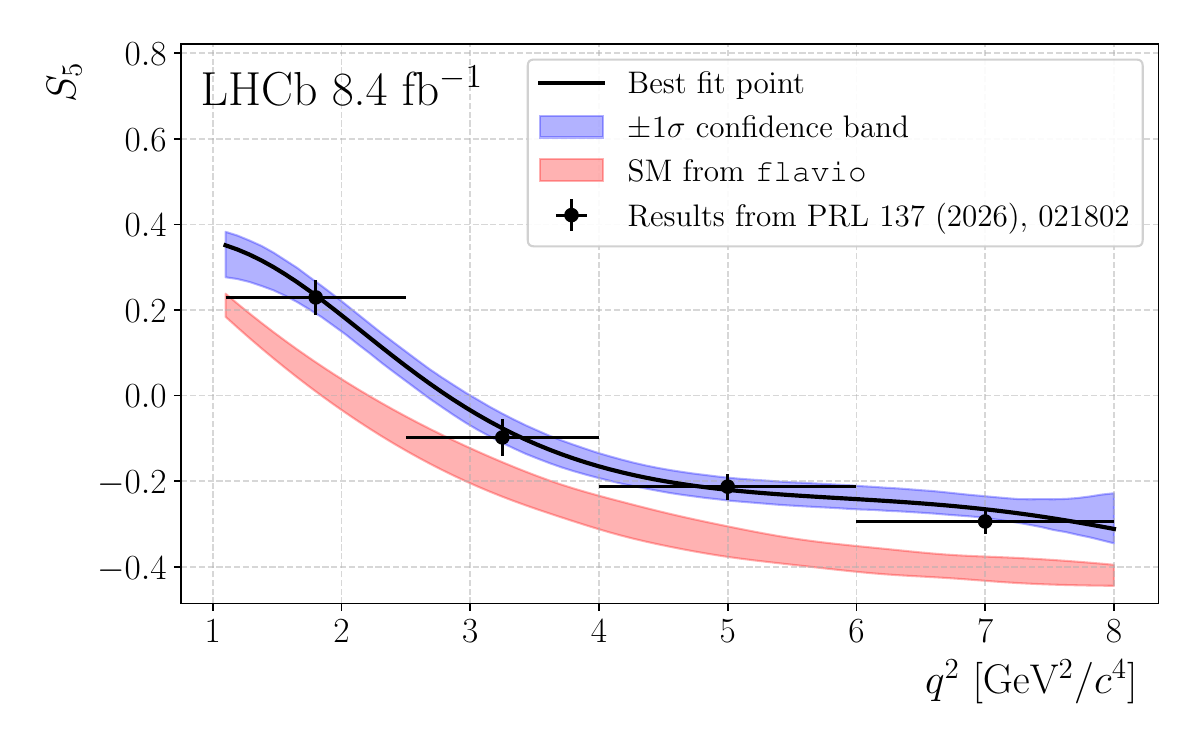}
    \includegraphics[width=0.49\linewidth]{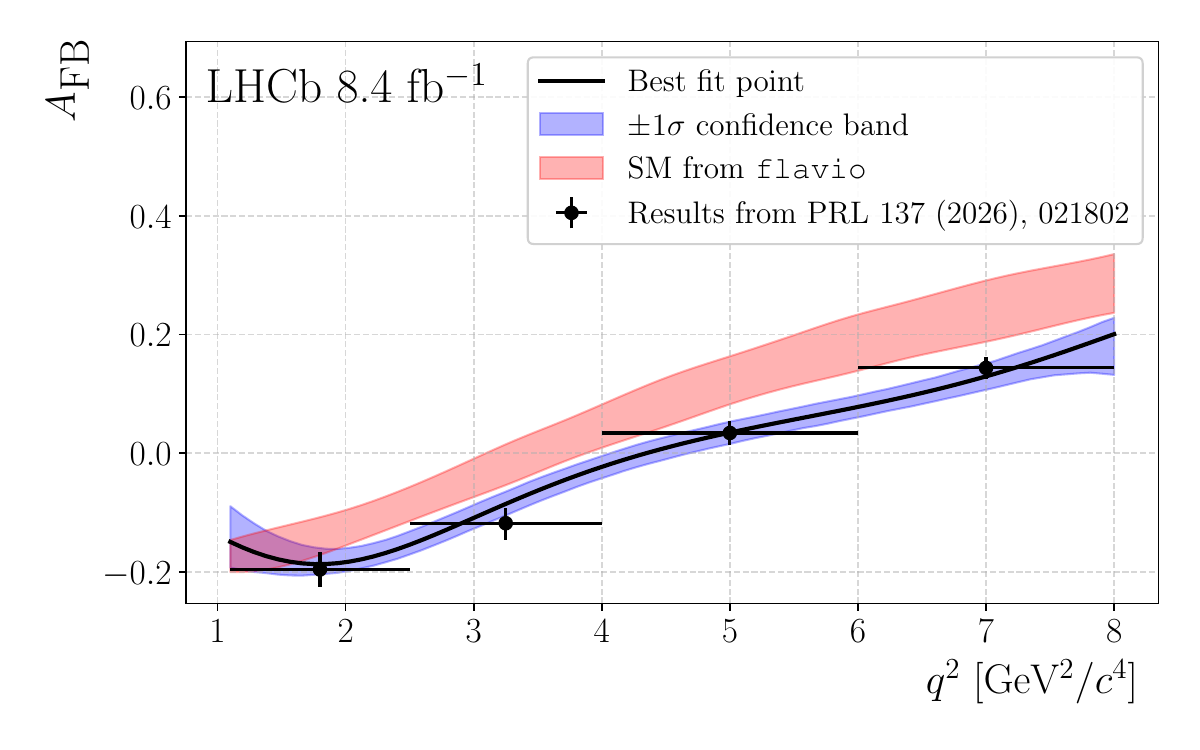}
    \caption{Angular observables $S_5$ and $A_\textrm{FB}$ as functions of $\mumu$ invariant mass squared, $q^2$. Shown are the (black line) best fit points with (blue) $\pm1\sigma$ bands obtained from sampling the total covariance matrix from the fit. The predictions obtained from \texttt{flavio}~\cite{straub:2018flavio} and results from the latest LHCb binned angular analysis~\cite{LHCb-PAPER-2025-041} are also shown.}
    \label{fig:fitresultscompare_main}
\end{figure}

The fit projections for the invariant mass $m_{\Kp\pim\mup\mun}$, the angular variables \ctl, \ctk, $\phi$, and \qsq are shown in Fig.~\ref{fig:fitprojections}. The results for the amplitudes are shown in Fig.~\ref{fig:fitresults1}. From the amplitude components, the amplitudes and angular observables are obtained by sampling from a multivariate bifurcated Gaussian distribution constructed from the correlation matrix fits to the one-dimensional log-likelihood profiles. The observables, defined in Eq.~\ref{eqn:Ss}, are computed from the amplitudes. The results for the observables are shown in Fig.~\ref{fig:fitresults2} together with the SM predictions obtained using the \texttt{flavio} package, where the default SM theory parameters and covariances are used. Some observables, in particular $S_5$ and $A_\text{FB}$, show disagreements with respect to the SM. The numerical results and uncertainties for the coefficients of the polynomial expansion of the transversity amplitudes (in the basis detailed in Eq.~\ref{eqn:ampbasis}) are given in Table~\ref{tab:results}.
The angular observables are also in good agreement with the most recent LHCb binned angular analysis~\cite{LHCb-PAPER-2025-041}. Figure~\ref{fig:fitresultscompare_main} shows $S_5$ and $A_\text{FB}$, with the results from Ref.~\cite{LHCb-PAPER-2025-041} overlaid.
All of the results, and covariances can be found on the HEPData repository \cite{HEPdata}.

\begin{table}[b]
\centering
\caption{Numerical results and uncertainties for the amplitude components. The results, the statistical correlation matrix and uncertainties, and the systematic covariance matrices are provided on the HEPData repository~\cite{HEPdata}.}
\renewcommand{\arraystretch}{1.3}
\begin{minipage}{0.48\textwidth}
\centering
\begin{tabular}{c c c}
\toprule
Amplitude & Coefficient & Value \\
\midrule
\multirow{8}{*}{$\mathcal{A}_\parallel^L$}
 & $\mathrm{Re}(\alpha_0)$ & $-0.071\;^{+0.006}_{-0.006}$ \\
 & $\mathrm{Re}(\alpha_1)$ & $-0.065\;^{+0.012}_{-0.011}$ \\
 & $\mathrm{Re}(\alpha_2)$ & $\phantom{-}0.019\;^{+0.013}_{-0.013}$ \\
 & $\mathrm{Re}(\alpha_3)$ & $\phantom{-}0.006\;^{+0.015}_{-0.014}$ \\
 & $\mathrm{Im}(\alpha_0)$ & $\phantom{-}0.013\;^{+0.004}_{-0.004}$ \\
 & $\mathrm{Im}(\alpha_1)$ & $\phantom{-}0.003\;^{+0.007}_{-0.007}$ \\
 & $\mathrm{Im}(\alpha_2)$ & $-0.003\;^{+0.009}_{-0.009}$ \\
 & $\mathrm{Im}(\alpha_3)$ & $\phantom{-}0.003\;^{+0.010}_{-0.010}$ \\
\midrule
\multirow{8}{*}{$\mathcal{A}_\parallel^R$} 
 & $\mathrm{Re}(\alpha_0)$ & $\phantom{-}0.051\;^{+0.008}_{-0.008}$ \vspace{0.025cm} \\
 & $\mathrm{Re}(\alpha_1)$ & $-0.042\;^{+0.015}_{-0.016}$ \vspace{0.025cm} \\
 & $\mathrm{Re}(\alpha_2)$ & $\phantom{-}0.028\;^{+0.016}_{-0.020}$ \vspace{0.025cm} \\
 & $\mathrm{Re}(\alpha_3)$ & $-0.029\;^{+0.018}_{-0.018}$ \vspace{0.025cm} \\
 & $\mathrm{Im}(\alpha_0)$ & $-0.011\;^{+0.014}_{-0.013}$ \vspace{0.025cm} \\
 & $\mathrm{Im}(\alpha_1)$ & $-0.043\;^{+0.028}_{-0.022}$ \vspace{0.025cm} \\
 & $\mathrm{Im}(\alpha_2)$ & $-0.027\;^{+0.030}_{-0.023}$ \vspace{0.025cm} \\
 & $\mathrm{Im}(\alpha_3)$ & $\phantom{-}0.012\;^{+0.033}_{-0.027}$ \vspace{0.025cm}  \\
\bottomrule
\end{tabular}
\end{minipage}\hfill\begin{minipage}{0.48\textwidth}
\centering
\begin{tabular}{c c c}
\toprule
Amplitude & Coefficient & Value \\
\midrule
\multirow{8}{*}{$\mathcal{A}_\perp^L$}
 & $\mathrm{Re}(\alpha_0)$ & $-0.022\;^{+0.004}_{-0.003}$ \\
 & $\mathrm{Re}(\alpha_1)$ & $-0.067\;^{+0.006}_{-0.006}$ \\
 & $\mathrm{Re}(\alpha_2)$ & $\phantom{-}0.032\;^{+0.008}_{-0.008}$ \\
 & $\mathrm{Re}(\alpha_3)$ & $-0.017\;^{+0.009}_{-0.009}$ \\
 & $\mathrm{Im}(\alpha_0)$ & $\phantom{-}0.014\;^{+0.007}_{-0.007}$ \\
 & $\mathrm{Im}(\alpha_1)$ & $\phantom{-}0.007\;^{+0.012}_{-0.012}$ \\
 & $\mathrm{Im}(\alpha_2)$ & $\phantom{-}0.024\;^{+0.014}_{-0.015}$ \\
 & $\mathrm{Im}(\alpha_3)$ & $\phantom{-}0.008\;^{+0.018}_{-0.018}$ \\
\midrule
\multirow{4}{*}{$\mathcal{A}_\perp^R$}
 & $\mathrm{Re}(\alpha_0)$ & $\phantom{-}0.066\;^{+0.006}_{-0.007}$ \\
 & $\mathrm{Re}(\alpha_1)$ & $-0.008\;^{+0.011}_{-0.012}$ \\
 & $\mathrm{Re}(\alpha_2)$ & $-0.015\;^{+0.013}_{-0.013}$ \\
 & $\mathrm{Re}(\alpha_3)$ & $\phantom{-}0.005\;^{+0.014}_{-0.015}$ \\
\midrule
\multirow{4}{*}{$\mathcal{A}_0^L$}
 & $\mathrm{Re}(\alpha_0)$ & $\phantom{-}0.191\;^{+0.003}_{-0.003}$ \\
 & $\mathrm{Re}(\alpha_1)$ & $-0.001\;^{+0.004}_{-0.004}$ \\
 & $\mathrm{Re}(\alpha_2)$ & $\phantom{-}0.004\;^{+0.006}_{-0.006}$ \\
 & $\mathrm{Re}(\alpha_3)$ & $\phantom{-}0.001\;^{+0.007}_{-0.007}$ \\
\bottomrule
\end{tabular}
\end{minipage}
\label{tab:results}
\end{table}

\begin{figure}
    \centering
    \includegraphics[width=0.7\linewidth]{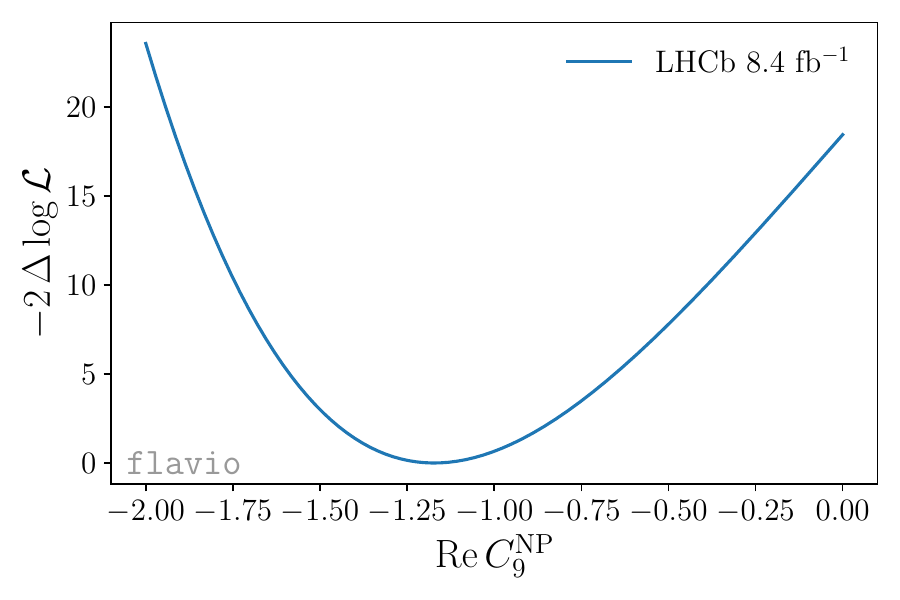}
    \caption{Profile likelihood as a function of $\textrm{Re}(C_9^{\textrm{NP}})$, obtained with \texttt{flavio}~\cite{straub:2018flavio} and constructing angular observables using the baseline binning scheme used in Ref.~\cite{LHCb-PAPER-2025-041}. Theoretical uncertainties associated with the nuisance parameters in \texttt{flavio} are included. All of the other Wilson coefficients are fixed to their SM values.}
    \label{fig:wilson1}
\end{figure}

\begin{figure}
    \centering
    \includegraphics[width=0.65\linewidth]{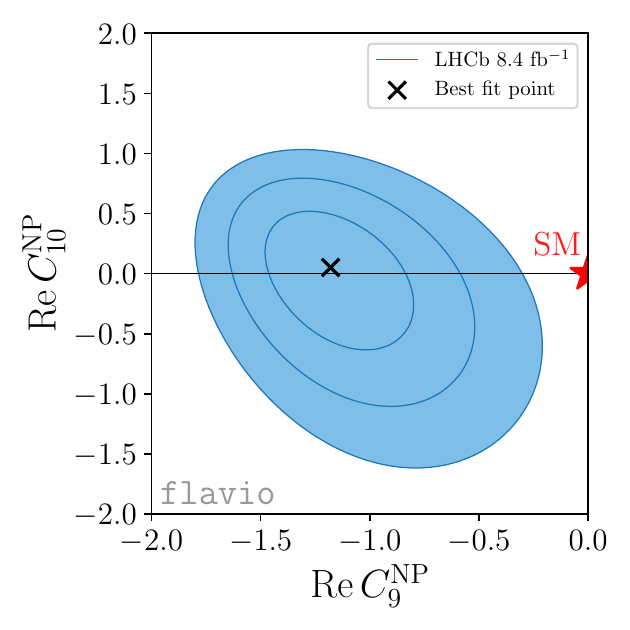}
    \caption{Profile likelihood as a function of $\textrm{Re}(C_9^{\textrm{NP}})$ and $\textrm{Re}(C_{10}^{\textrm{NP}})$, obtained with \texttt{flavio}~\cite{straub:2018flavio} and constructing angular observables using the baseline binning scheme used in Ref.~\cite{LHCb-PAPER-2025-041}. The $1\sigma$, $2\sigma$, and $3\sigma$ contours are shown. Theoretical uncertainties associated with the nuisance parameters in \texttt{flavio} are included. All of the other Wilson coefficients are fixed to their SM values. The SM value is at ($\textrm{Re}(C_9^{\textrm{NP}})$, $\textrm{Re}(C_{10}^{\textrm{NP}})$) = ($0.0$, $0.0$).}
    \label{fig:wilson2}
\end{figure}

To quantify the level of agreement with the SM, a log-likelihood profile as a function of the Wilson coefficient $\textrm{Re}(C_9)$ ascribed to NP, $\textrm{Re}(C_9^{\textrm{NP}})\equiv \textrm{Re}(C_9)-\textrm{Re}(C_9^{\textrm{SM}})$, is made using the \texttt{flavio} package. The profile is obtained by constructing angular observables applying the baseline binning scheme used in Ref.~\cite{LHCb-PAPER-2025-041}. Theoretical uncertainties associated with nuisance parameters are included through the covariance matrix obtained from the \texttt{FastLikelihood} approach~\cite{straub:2018flavio}. All of the other Wilson coefficients are fixed to their SM values. The best-fit value is ${\textrm{Re}(C_9^{\textrm{NP}})=-1.17^{+0.21}_{-0.20}}$. Based on Wilks' theorem~\cite{Wilks:1938dza}, the deviation from the SM expectation corresponds to a significance of $4.3\sigma$. Results from the profiles in $\textrm{Re}(C_9^{\textrm{NP}})$ and the $\textrm{Re}(C_9^{\textrm{NP}})/\textrm{Re}(C_{10}^{\textrm{NP}})$ plane are shown in Figs.~\ref{fig:wilson1} and \ref{fig:wilson2}. If narrower $\qsq$ bins are used, in this case bins which are a quarter of the size of the bins used in Ref.~\cite{LHCb-PAPER-2025-041}, the value obtained is ${\textrm{Re}(C_9^{\textrm{NP}})=-1.17^{+0.19}_{-0.18}}$. The deviation with respect to the SM expectation increases to $4.8\sigma$, demonstrating that increased resolution in $\qsq$ enhances sensitivity to potential NP contributions.

\section{Conclusions}
\label{sec:conclusions}

The amplitudes of the decay \BdToKstmm have been measured for the first time in a model-independent way using Legendre polynomials in the region \qsqregion. This analysis enhances the interpretability of this decay process, allowing for a precise determination of potential NP contributions and test the size of potential unaccounted hadronic effects. 
The \CP-averaged observables show disagreements with a set of SM predictions obtained from Ref.~\cite{straub:2018flavio} as seen in Fig.~\ref{fig:fitresults2}. In addition, interpreting these results through the \texttt{flavio} framework, the results present strong deviations from the SM predictions, at the level of $4.3\sigma$ to $4.8\sigma$.

As theoretical predictions develop and improve and larger datasets become available, this approach will provide a deep insight into the origin of the anomalies. A larger dataset will also allow for a more precise measurement of the S-wave, and additional quantities such as the \CP-asymmetries and scalar amplitudes.

%% file: LHCb/acknowledgements.tex
\section*{Acknowledgements}
%
%
\noindent We express our gratitude to our colleagues in the CERN
accelerator departments for the excellent performance of the LHC. We
thank the technical and administrative staff at the LHCb
institutes.
We acknowledge support from CERN and from the national agencies:
ARC (Australia);
CAPES, CNPq, FAPERJ and FINEP (Brazil); 
MOST and NSFC (China); 
CNRS/IN2P3 and CEA (France);  
BMFTR, DFG and MPG (Germany);
NKFIH (Hungary);              
INFN (Italy); 
NWO (Netherlands); 
MNiSW and NCN (Poland); 
MEC/IFA (Romania); 
MICIU and AEI (Spain);
SNSF and SER (Switzerland); 
NASU (Ukraine); 
STFC (United Kingdom); 
DOE NP and NSF (USA).
We acknowledge the computing resources that are provided by ARDC (Australia), 
CBPF (Brazil),
CERN, 
IHEP and LZU (China),
IN2P3 (France), 
KIT and DESY (Germany), 
INFN (Italy), 
SURF (Netherlands),
Polish WLCG (Poland),
IFIN-HH (Romania), 
PIC (Spain), CSCS (Switzerland), 
GridPP (United Kingdom),
and NSF (USA).  
We are indebted to the communities behind the multiple open-source
software packages on which we depend.
Individual groups or members have received support from
RTP (Australia), 
FWO Odysseus grant G0ASD25N (Belgium), 
Key Research Program of Frontier Sciences of CAS, CAS PIFI, CAS CCEPP (China); 
Minciencias (Colombia);
EPLANET, Marie Sk\l{}odowska-Curie Actions, ERC and NextGenerationEU (European Union);
A*MIDEX, ANR, IPhU and Labex P2IO, and R\'{e}gion Auvergne-Rh\^{o}ne-Alpes (France);
Alexander-von-Humboldt Foundation (Germany);
ICSC (Italy); 
Severo Ochoa and Mar\'ia de Maeztu Units of Excellence, GVA, XuntaGal, GENCAT, InTalent-Inditex and Prog.~Atracci\'on Talento CM (Spain);
the Leverhulme Trust, the Royal Society and UKRI (United Kingdom).

%% file: Authorship_LHCb-PAPER-2026-023.tex
\centerline
{\large\bf LHCb collaboration}
\begin
{flushleft}
\small
R.~Aaij$^{39}$\lhcborcid{0000-0003-0533-1952},
M.~Abdelfatah$^{71}$,
A.S.W.~Abdelmotteleb$^{59}$\lhcborcid{0000-0001-7905-0542},
C.~Abellan~Beteta$^{53}$\lhcborcid{0009-0009-0869-6798},
F.~Abudin\'en$^{61}$\lhcborcid{0000-0002-6737-3528},
T.~Ackernley$^{63}$\lhcborcid{0000-0002-5951-3498},
A.A.~Adefisoye$^{71}$\lhcborcid{0000-0003-2448-1550},
B.~Adeva$^{49}$\lhcborcid{0000-0001-9756-3712},
M.~Adinolfi$^{57}$\lhcborcid{0000-0002-1326-1264},
P.~Adlarson$^{87,44}$\lhcborcid{0000-0001-6280-3851}
% ...

{\footnotesize \it

% ...
$^{39}$Nikhef National Institute for Subatomic Physics, Amsterdam, Netherlands\\
% ...
$^{44}$National Center for Nuclear Research (NCBJ), Warsaw, Poland\\
% ...
$^{49}$Instituto Galego de F{\'\i}sica de Altas Enerx{\'\i}as (IGFAE), Universidade de Santiago de Compostela, Santiago de Compostela, Spain\\
% ...
$^{53}$Physik-Institut, Universit{\"a}t Z{\"u}rich, Z{\"u}rich, Switzerland\\
% ...
$^{57}$H.H. Wills Physics Laboratory, University of Bristol, Bristol, United Kingdom\\
% ...
$^{59}$Department of Physics, University of Warwick, Coventry, United Kingdom\\
% ...
$^{61}$School of Physics and Astronomy, University of Edinburgh, Edinburgh, United Kingdom\\
$^{62}$School of Physics and Astronomy, University of Glasgow, Glasgow, United Kingdom\\
$^{63}$Oliver Lodge Laboratory, University of Liverpool, Liverpool, United Kingdom\\
% ...
$^{71}$Syracuse University, Syracuse, NY, United States\\
% ...
$^{87}$Department of Physics and Astronomy, Uppsala University, Uppsala, Sweden, associated to $^{62}$\\
% ...
}
\end{flushleft}